\documentclass[journal,10pt]{IEEEtran}
\usepackage{amsfonts}
\usepackage{mathrsfs}
\usepackage{amssymb}
\usepackage{wrapfig}
\usepackage{amsmath}
\usepackage{accents}
\usepackage{dsfont}
\usepackage{amsbsy}
\usepackage{setspace}
\usepackage{graphicx}
\usepackage{subfig}
\usepackage{url}
\usepackage{color}
\usepackage{epstopdf}
\usepackage{amssymb}
\usepackage[top=0.75in, bottom=1in, left=0.625in, right=0.625in, textwidth=7.25in,textheight=9.25in]{geometry}

\def\argmax{\operatornamewithlimits{arg\,max}}

\IEEEoverridecommandlockouts

\begin{document}
\title{Throughput-Optimal Scheduling Design with Regular Service Guarantees in Wireless Networks
\author{Bin Li, Ruogu Li, and Atilla Eryilmaz}\thanks{Bin Li (celibin@gmail.com), Ruogu Li (lirg03@gmail.com), and Atilla Eryilmaz
(eryilmaz.2@osu.edu) are with the Department of Electrical
and Computer Engineering at The Ohio State University, Columbus,
Ohio 43210 USA.}}
\date{}

\maketitle
\newtheorem{theorem}{Theorem}
\newtheorem{lemma}{Lemma}
\newtheorem{claim}{Claim}
\newtheorem{proposition}{Proposition}
\newtheorem{corollary}{Corollary}
\newtheorem{definition}{Definition}
\newtheorem{assumption}{Assumption}
\newtheorem{remarks}{Remarks}

\newcommand{\cS}{\mathcal{S}}
\newcommand{\vS}{\mathbf{S}}
\newcommand{\vQ}{\mathbf{Q}}
\newcommand{\bE}{\mathds{E}}
\newcommand{\mc}{\mathcal}
\newcommand{\mb}{\mathbf}
\newcommand{\bs}{\boldsymbol}
\newcommand{\ol}{\overline}

\makeatletter
\newcommand{\rmnum}[1]{\romannumeral #1}
\newcommand{\Rmnum}[1]{\expandafter\@slowromancap\romannumeral #1@}
\makeatother
\def\maximize{\operatornamewithlimits{Maximize}}
\def\limitsup{\operatornamewithlimits{limsup}}
\def\limitinf{\operatornamewithlimits{liminf}}

\begin{abstract}
Motivated by the regular service requirements of video applications for improving Quality-of-Experience (QoE) of users, we consider the design of scheduling strategies in multi-hop wireless networks that not only \emph{maximize system throughput } but also \emph{provide regular inter-service times} for all links. Since the service regularity of links is related to the higher-order statistics of the arrival process and the policy operation, it is highly challenging to characterize and analyze directly. We overcome this obstacle by introducing a new quantity, namely the \emph{time-since-last-service} (TSLS), which tracks the time since the last service. By combining it with the queue-length in the weight, we propose a novel maximum-weight type scheduling policy, called Regular Service Guarantee (RSG) Algorithm. The unique evolution of the TSLS counter poses significant challenges for the analysis of the RSG Algorithm. 

To tackle these challenges, we first propose a novel Lyapunov function to show the throughput optimality of the RSG Algorithm. Then, we prove that the RSG Algorithm can provide service regularity guarantees by using the Lyapunov-drift based analysis of the steady-state behavior of the stochastic processes. In particular, our algorithm can achieve a degree of service regularity within a factor of a fundamental lower bound we derive. This factor is a function of the system statistics and design parameters and can be as low as two in some special networks. Our results, both analytical and numerical, exhibit significant service regularity improvements over the traditional throughput-optimal policies, which reveals the importance of incorporating the metric of time-since-last-service into the scheduling policy for providing regulated service.

\end{abstract}

\section{Introduction}

During the past years, there has been increasing deployment of a variety of real-time applications over the wireless networks, especially streaming multi-media applications. Unlike its non-real-time counterpart, the real-time traffic often has various quality-of-service (QoS) requirements besides throughput. Such requirements usually include end-to-end delay constraints, packet delivery ratio requirements, and the regularity of the inter-service times. Unlike the traditional long-term mean throughput based requirements, these QoS requirements often have a complex dependence on the higher-order statistics of the arrival process as well as the system operation. Thus, the canonical optimization-based approaches that aim to optimize the throughput performance (e.g., \cite{taseph92,erysri05,linshr04,neemodli05,linshrsri06}) do not apply.

Recently, valuable efforts have been exerted in the design of algorithms that improve various aspects of the QoS, especially on the delay performance of the algorithms. For example, some works focus on designing algorithms with low end-to-end delay performance, such as \cite{buisristo09,yinshared09,xiolieryeki11}. Constant delay bounds (e.g. \cite{nee10}) and delivery ratio requirements for deadline-constrained traffic (e.g. \cite{houborkum09,houkum09,houkum10,jarsri10,lierytwc13}) are some of the other QoS metrics considered in the literature.

\begin{figure}[htb!]
\begin{center}
\includegraphics[scale=0.5]{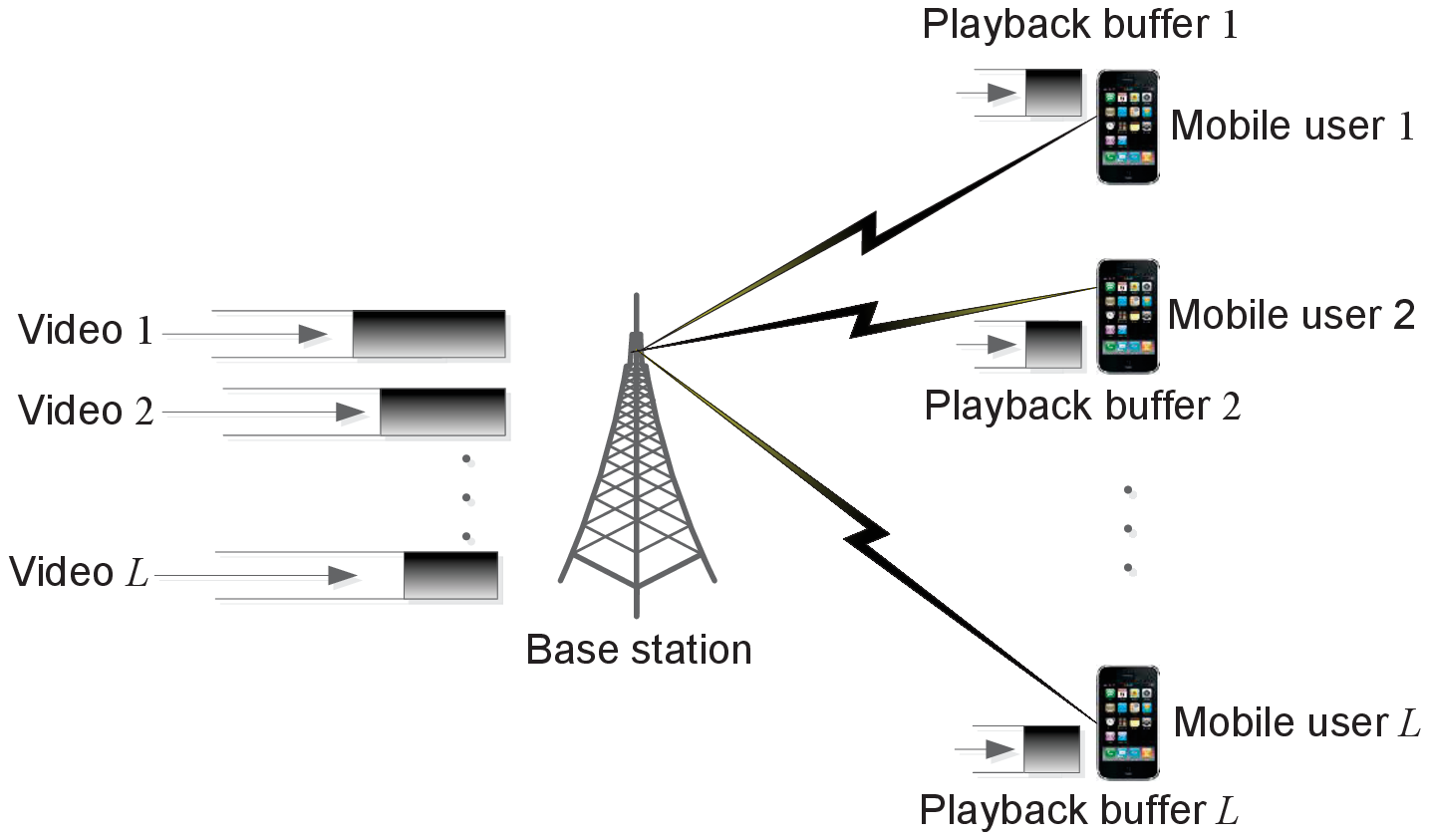}
\caption{Cellular network with a single base station and $L$ users}
\label{fig:downlink}
\vspace{-0.2in}
\end{center}
\end{figure}

However, these QoS metrics do not fully characterize the Quality-of-Experience (QoE) of users in video applications in wireless networks. To see it, we can envision the network scenario where each individual user wants to download its video from the base station, as shown in Fig. \ref{fig:downlink}.
Each mobile user would like to receive the data from the base station regularly. Indeed, the QoE of users is highly related to the average Perceived Video Quality (PVQ) across the sequence of scenes forming the video, where the PVQ traditionally is a local quality measure associated with a particular scene or a short period of time. In \cite{yimbovik09, joseph09}, the authors point out that the variance in PVQ leads to the worse QoE than the constant quality video with even smaller average PVQ. Yet, both the time-varying nature of wireless channels and the scheduling policy significantly affect the variance of the received data of each mobile user. Traditional scheduling policies aiming to maximize the system throughput or minimize the delay at the base station side do not take users' experience into account and thus lead to the high variance of the received data of mobile users. 



This motivates us to reduce variability of arrivals to the mobile users, which can be achieved by providing regulated inter-service times for the arriving flows at the base station end.
However, the inter-service time characteristics are difficult to analyze directly due to: its complex dependence on the high-order statistics of the arrival and service processes, and its non-Markovian evolution. To overcome this, we need to find new approaches to study the inter-service time behavior. To the best of our knowledge, this is the first work that rigorously studies the service regularity of the scheduling policies. Our contributions in this work can be summarized as follows:

\noindent $\bullet$ We introduce a new quantity (cf. Section~\ref{sec:model}), namely the \emph{time-since-last-service,} that has a tight relationship with the service regularity performance, and hence enables novel design strategies. Yet, this new parameter has its unique evolution, drastically different from a queue, which poses new challenges for its  analysis.

\noindent $\bullet$ We develop a novel maximum-weight type scheduling policy that combines the time-since-last-service parameter and the queue-length in its weight measure (cf. Section~\ref{sec:alg}). Then, we show that the proposed scheduling policy possesses the desirable throughput optimality property by using a novel Lyapunov function.

\noindent $\bullet$ We derive lower and upper bounds on the service regularity performance (cf. Section~\ref{sec:regular}) by utilizing a novel Lyapunov-drift-based argument, inspired by the approach in \cite{erysri12}. We further show that, by properly scaling the design parameter in our policy, we can guarantee a degree of service regularity within a factor of our fundamental lower bound. This factor is a function of the system statistics and design parameters and can be as low as two under symmetric arrival rates in some special networks. 

\noindent $\bullet$ We support our analytical results with extensive numerical investigations (cf. Section~\ref{sec:simulation}), which show significant performance gains in the service regularity over the traditional queue-length-based policies. Furthermore, the numerical investigations indicate that the service regularity performance of our policy actually  approaches the lower bounds as the weight of the time-since-last-service increases in some special networks.

This work extends our earlier work \cite{lieryilmazinfocom13} in several key aspects: (1) we 
conduct novel analyses that extend both throughput optimality and service regularity guarantee results to general multi-hop fading networks; (2) we show the existence of all moments of the system state under our proposed algorithm, which establishes the foundation to utilize the Lyapunov-drift based analysis of the steady-state behavior of stochastic processes; (3) we conduct simulations to compare our policy with traditional queue-length-based scheduling algorithms in more general setups, including switch topologies and fading scenarios.



\section{system model}
\label{sec:model}

We consider a wireless network with $L$ links, where a link represents a pair of
a transmitter and a receiver that are within the transmission range of each other.
We assume that the system operates in slotted time with normalized slots 
$t\in\{1,2,...\}$. Due to the interference-limited nature of wireless transmissions, the
success or failure of a transmission over a link depends on whether an
interfering link is also active in the same slot, which is
called the \emph{link-based conflict model}.
We call a set of links that can be active simultaneously as a
\emph{feasible schedule} and
denote it as $\mb{S}[t]=(S_l[t])_{l=1}^{L}$, where
$S_l[t]=1$ if the link $l$ is scheduled in slot $t$ and $S_l[t]=0$, otherwise. We use $\mc{S}$ to denote the set of all feasible schedules.

We capture the channel fading over link $l$ via a non-negative-integer-valued
random variable $C_l[t]$, with $C_l[t]\leq C_{\max}$, $\forall l,t$, for some
$C_{\max}<\infty$, which measures the maximum amount of service
available in slot $t$, if the link $l$ is scheduled. We assume that $\mb{C}[t]=(C_l[t])_{l=1}^{L}$, $\forall t\geq0$, 
are independently and identically distributed (i.i.d.) over time. We assume that $\ol{c}_{\min}\triangleq\min_{l}\bE[C_l[t]]>0$. Let $\mc{S}^{(\mb{c})}\triangleq\{\mb{S}\mb{c}:\mb{S}\in\mc{S}\}$ denote the set of feasible rate vectors when the channel is in state $\mb{c}$, where 
$\mb{a}\mb{b}=(a_lb_l)_{l=1}^{L}$ denotes the component-wise product of two vectors $\mb{a}$ and $\mb{b}$. 
Then, the \emph{capacity region} is defined as
\begin{align}
\mc{R}\triangleq\sum_{\mb{c}}\Pr\{\mb{C}[t]=\mb{c}\}\cdot\text{CH}\{\mc{S}^{(\mb{c})}\},
\end{align}
where $\text{CH}\{\mc{A}\}$ denotes a convex hull of the set $\mc{A}$, and the summation is a Minkowski addition of sets.

We assume a per-link traffic model\footnote{We note that our algorithm can be extended to serve multi-hop traffic, but the notion of service regularity is clearer in the per-link context.}, where $A_l[t]$ denotes the number of
packets arriving at link $l$ in slot $t$
that are independently distributed over links,
and i.i.d. over time with finite mean $\lambda_l>0$,
and $A_l[t]\leq A_{\max}$, $\forall l,t,$ for some $A_{\max}<\infty$. 
Accordingly, a queue is maintained for each link $l$ with $Q_l[t]$ denoting its
queue length at the beginning of time slot $t$.
Then, the evolution of queue $l$ is
described as follows:
\begin{align}
Q_l[t+1]=(Q_l[t]+A_l[t]-C_l[t]S_l[t])^{+}, \forall l,
\end{align}
where $(x)^{+}=\max\{x,0\}$. We say that the queue $l$ is \emph{strongly stable}
if it satisfies
\begin{align}
\limsup_{T\rightarrow\infty}\frac{1}{T}\sum_{t=1}^{T}\bE[Q_l[t]]<\infty.
\end{align}
We call system \emph{stable} if all queues are strongly stable. In this paper, we consider the policies under which the system evolves as a Markov Chain. We call an algorithm \emph{throughput-optimal} if it makes all queues strongly
stable for any arrival rate vector $\bs{\lambda}=(\lambda_l)_{l=1}^{L}$ that lies
strictly within the capacity region.

In this work, we are interested in providing regular service for each link, which
relates to the statistics of the \emph{inter-service time}.
We use $I_l[m]$ to denote the time between the $(m-1)^{th}$ and the $m^{th}$
service for link $l$. If the system is stable,
the \emph{steady-state distribution} of the underlying Markov Chain
exists (see \cite{neely10bk}) and thus we use $\ol{\mb{Q}}=(\ol{Q}_l)_{l=1}^{L}$, 
$\ol{\mb{S}}=(\ol{S}_l)_{l=1}^{L}$ and $\ol{\mb{I}}=(\ol{I}_l)_{l=1}^{L}$ to denote the random vector
with the same steady-state distribution of the queue-length, service processes and inter-service time, respectively. We use the normalized 
second moment of the inter-service time under the steady-state
distribution, i.e., $\bE[\ol{I}_l^2]/(\bE[\ol{I}_l])^2$, as a measure of
the ``regularity'' of the service that link $l$ receives. Noting that $\bE[\ol{I}_l^2]/(\bE[\ol{I}_l])^2=\text{Var}(\ol{I}_l)/(\bE[\ol{I}_l])^2+1$, the normalized second moment of the inter-service time reflects its normalized variance. Hence, 
the smaller the normalized second moment of the inter-service time, the smaller its normalized variance and thus the received service is more regular.
%


We would like to develop throughput-optimal policies
that achieve low values of a linear increasing function of $(\bE[\ol{I}_l^2]/(\bE[\ol{I}_l])^2)_{l=1}^{L}$ 
in steady-state, implying more regular service. However, unlike queue-lengths with Markovian evolution,
the dynamics of inter-service times do not lend themselves to commonly used
Markovian analysis methods. To overcome this obstacle, we introduce the
following related quantity, namely the \emph{time-since-last-service}, which has
much more tractable form of evolution, and whose mean has a
close relationship to the normalized second moment of the inter-service time (cf. Lemma \ref{lmm:TandI}).

For each link $l$, we introduce a counter $T_l$, namely Time-Since-Last-Service
(TSLS), to keep track of the time since it was lastly \emph{served},
i.e., it was scheduled and the channel was available. Let
\begin{eqnarray*}
\tau_l[t] \triangleq \max_{\tau=\{1,\dots,t-1\}} \left\{
\begin{array}{l}
S_l[\tau]C_l[\tau]>0, S_l[\tau+1]C_l[\tau+1]=\\
\dots=S_l[t-1]C_l[t-1]=0
\end{array}\right\},
\end{eqnarray*}
be the last time when link $l$ was served before time slot $t$, then
$T_l[t]=t-\tau_l[t]-1$. By definition, each counter $T_l$ increases by 1 in
each time slot when link $l$ has zero transmission rate, either because it is not scheduled, or because its channel is unavailable, i.e., $C_l[t]=0$, and drops to 0, otherwise. 
More precisely, the evolution of the counter $T_l$ can be written as
\begin{equation}
\label{eqn:TLSL:evolution}
T_l[t+1] = \left\{ \begin{array}{ll}
         0     & \mbox{if $S_l[t]C_l[t]>0$};\\
      T_l[t]+1 & \mbox{if $S_l[t]C_l[t]=0$}.\end{array} \right.
\end{equation}

\begin{figure}[htp]
\vspace{-0.1in}
\centering \resizebox{200pt}{!}{\includegraphics{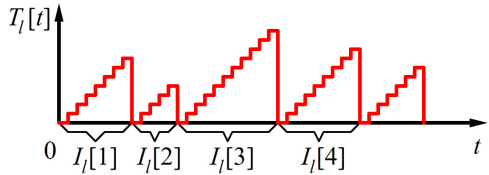}}
\caption{A sample trajectory of $I_l[m]$ and $T_l[t]$, where the curve shows the evolution of $T_l[t]$.}\vspace{-0.1in}\label{fig:sampleTandI}
\end{figure}

It can be seen from \eqref{eqn:TLSL:evolution} that the evolution of
$T_l[t]$ differs significantly from that of a traditional queue
(also see Fig. \ref{fig:sampleTandI}). In particular, unlike the
slowly evolving nature of queue-lengths, the $T_l[t]$ is incremented until
link $l$ receives service at which time it drops to zero. In our design,
we will consider policies that not only use queue-lengths to achieve
throughput-optimality, but also include TSLS to improve service regularity. 

The evolution of $T_l$ is tightly related to the inter-service time $I_l$, where
$I_l$ is the time between two consecutive instances when $T_l$ hits zero, as shown in 
Fig. \ref{fig:sampleTandI}. In fact, we have the following lemma relating
the two in steady-state.

\begin{lemma}
\label{lmm:TandI}
For any policy under which the steady-state distribution of the underlying Markov Chain exists, 
we have
\begin{eqnarray}
\bE[\overline{T}_l]=\frac{1}{2} \left( \frac{1}{\bE[\overline{I}_l]}\bE[\overline{I}_l^2] - 1 \right),
\label{eqn:TandI}
\end{eqnarray}
where $\ol{T}_l$ and $\ol{I}_l$ denote the steady-state
TLSL and inter-service time at link $l$, respectively. 
\end{lemma}

\begin{IEEEproof}
The detailed proof is provided in Appendix~\ref{app:proofLmmTandI}.
\end{IEEEproof}

Lemma~\ref{lmm:TandI} reveals the connection between the second moment
of the inter-service time $\overline{I}_l$ and the mean of TLSL
$\overline{T}_l$ in steady-state.
This can be intuitively seen in Fig.~\ref{fig:sampleTandI},
where the area of each ``triangle'' under the trajectory of $T_l[t]$
is roughly $\frac{1}{2}I_l^2$. In this work, we are interested in designing throughput-optimal algorithms that reduce the total weighted-sum\footnote{The weighting parameter $\rho_l$ is the arrival intensity at link $l$, and indicates that  the link with higher load prefers more regular service. Despite this, $\rho_l$ is included in the objective function primarily for technical reasons. Noting that the link preference parameter $\beta_l$ can be any non-negative real number, and thus this weighted form is still general enough. } of the normalized second moment of the inter-service time, i.e., $\sum_{l=1}^{L}\beta_l\rho_l\bE[\ol{I}_l^2]/\left(\bE[\ol{I}_l]\right)^2$, where $\mu_l\triangleq1/\bE[\ol{I}_l]$, $\rho_l\triangleq\lambda_l/\mu_l$, and $\beta_l\geq0$ is some parameter related to the link. We can set $\beta_l>0$ if link $l$ prefers regular service and $\beta_l=0$ otherwise.

According to Lemma \ref{lmm:TandI}, we have
\begin{align}
\sum_{l=1}^{L}\beta_l\rho_l\frac{\bE[\ol{I}_l^2]}{\left(\bE[\ol{I}_l]\right)^2}=2\sum_{l=1}^{L}\beta_l\lambda_l\bE[\ol{T}_l]+\sum_{l=1}^{L}\beta_l\lambda_l.
\end{align}

Since $\sum_{l=1}^{L}\beta_l\lambda_l$ only depends on the system parameters, 
we will use 
$\sum_{l=1}^{L}\beta_l\lambda_l\bE[\ol{T}_l]$ as our measure for the service regularity. In this work, we aim to design a scheduling policy that
is not only throughput-optimal, but also yields provable good characteristics
in the service regularity.

We achieve this dual objective by developing a parametric class of throughput-optimal schedulers
(cf. Section \ref{sec:policy}) that utilize a combination of queue-lengths and TSLS in its
decisions. Our policy is shown to guarantee a ratio (as a function of 
the system statistics) in its
service regularity with respect to a fundamental lower bound (cf. Section \ref{sec:lb}).

\section{Algorithm design for regular service}
\label{sec:alg}
In this section, we first discuss the
inefficiency of the well-known throughput-optimal Maximum Weight Scheduling (MWS)
Algorithm in
terms of service regularity. We then propose
Regular Service Guarantee policy which can be shown that not only achieves the
throughput optimality but also possesses good service regularity performance.

\subsection{Inefficiency of the MWS Algorithm}
\label{sec:MWS}


In this subsection, we describe a well-known scheduling policy, namely
the Maximum Weight Scheduling (MWS) Algorithm and discuss its inefficiency
in terms of service regularity performance. We first give the definition
of the MWS Algorithm for completeness.

\begin{definition}[Maximum Weight Scheduling (MWS) Algorithm] \label{def:MW}
Under our model, the MWS Algorithm selects a schedule
$\mb{S}^{(\text{MWS})}[t]$ with the largest total sum of the product of queue-length and the maximum channel available rate within that schedule, i.e., it
chooses
\begin{eqnarray}
\mb{S}^{(\text{MWS})}[t] \in \argmax_{\mb{S}\in\mc{S}}\sum_{l=1}^{L}Q_l[t]C_l[t]S_l[t].
\end{eqnarray}
\end{definition}

The MWS Algorithm is known to be throughput-optimal
(e.g., \cite{taseph92,linshrsri06,neemodroh03,erysri06a}),
i.e., it stabilizes the network for any arrival rate vector
$\bs{\lambda}$ that strictly lies within the capacity region $\mc{R}$. In our setup,
the MWS Algorithm can be expected to have close-to-lower-bound average
delay performance (see \cite{gupshr09}). It has also been shown to be heavy-traffic optimal (see \cite{sto04,erysri12}), i.e., it minimizes the mean steady-state queue-length
under heavy-traffic conditions, where the arrival rate vector approaches the boundary of the capacity region from below.

However, despite its throughput optimality and a number of
favorable properties on the delay performance,
the MWS Algorithm may result in poor performance
in terms of service regularity. This can be
observed when the MWS Algorithm serves a set
of links with heterogeneous arrival statistics in a non-fading single-hop
network with uniform link rates. 

\begin{figure}[htp!]
\centering {\includegraphics[scale=0.4]{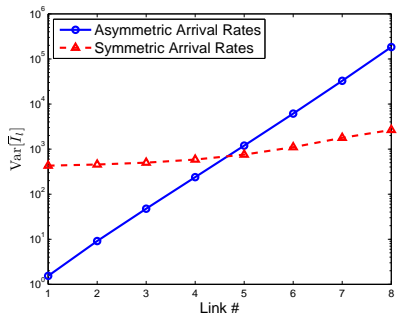}}
\caption{The variance of the inter-service time under the MWS Algorithm
for links with different arrival processes.
The links with smaller rates or more bursty arrivals suffer
from high variance of the inter-service time.}\label{fig:MWBad}
\vspace{-0.2in}
\end{figure}


In Fig.~\ref{fig:MWBad}, the blue line shows a scenario
where the $l^{th}$ link has a Bernoulli arrival with rate
$2^{-l}$ for $l\in \{1,\dots,8\}$. In this case, we observe
that the variance of the inter-service time increases
exponentially as the arrival rate of the link reduces.
The red curve illustrates a different scenario where all
$8$ links have the same mean arrival rate, but increasing variances
(i.e., burstiness) in their arrivals, where we observe
that the link with more bursty arrivals suffers from
higher variance in its inter-service time.




\subsection{The Regular Service Guarantee Policy}
\label{sec:policy}
As discussed above, the MWS Algorithm is throughput-optimal but
inefficient in providing regular services. Note that the introduced
TLSL counter has a direct impact on service regularity: the smaller
the mean TLSL value, the more regular the service. This interesting
connection motivates the following parametrized policy which is
later revealed to possess the characteristics of throughput
optimality and service regularity.

%

\vspace{0.1in}\hrule\vspace{0.1in}
\begin{definition}[Regular Service Guarantee (RSG) Algorithm]\label{defn:QPlusTPolicy}
In each time slot $t$, select a schedule $\mb{S}^{*}[t]$ such that
\begin{align}
\mb{S}^{*}[t]\in\argmax_{\mb{S}\in\mc{S}}\sum_{l=1}^{L}(\alpha_lQ_l[t]+\gamma\beta_l T_l[t])C_l[t]S_l,
\end{align}
where $\alpha_l>0$ and $\gamma\geq0$ are fixed control
parameters.
\end{definition}
\vspace{0.1in}\hrule\vspace{0.1in}

We note that there are two sets of control parameters in the
RSG Algorithm\footnote{The RSG Algorithm inherits the same complexity issue as the well-known MWS Algorithm. The low complexity or the distributed implementations of the RSG Algorithm are always attractive in practical networks and are left for future research.} and they affect different behaviors of the algorithm.
Yet, it will be revealed later that none of them affects its
throughput optimality. The parameters $\alpha_l$ are weighing factors for the queue-lengths,
where a larger $\alpha_l$ will result in a smaller average queue-length.
The parameter $\gamma$ is a common weighing factor of TSLS for all links. It will be revealed in Section \ref{sec:ub} that the design parameter
$\gamma$ can improve the service regularity 
as it increases. Also note that when $\gamma=0$, our policy coincides with the
MWS Algorithm. When $\gamma>0$, with the addition of $T_l[t]$
terms in the weight of each link, our algorithm operates completely different
from the MWS and its approximate algorithms, which, to the best of our knowledge,
are the only known policies possessing
the throughput-optimality characteristic in general
multi-hop network topologies. Despite of this, we can still show
that our algorithm is throughput-optimal.
\begin{proposition}
\label{prop:throughput}
The RSG Algorithm with any $\alpha_l>0$ and $\gamma\geq0$,
is throughput-optimal, i.e., for
any arrival rate vector $\bs{\lambda}\in\text{Int}(\mc{R})$,
the RSG Algorithm stabilizes the system, with
\begin{align}
\limsup_{K\rightarrow\infty} \frac{1}{K}\sum_{t=0}^{K-1}\sum_{l=1}^{L}\alpha_l\bE[Q_{l}[t]] \leq \frac{B(\boldsymbol{\alpha},\boldsymbol{\beta},\gamma)}{2\epsilon},
\end{align}
where $\text{Int}(\mc{A})$ denotes the interior points of the region $\mc{A}$,
$B(\boldsymbol{\alpha},\boldsymbol{\beta},\gamma)\triangleq 4\gamma C_{\max}\sum_{l=1}^{L}\beta_l+\sum_{l=1}^{L}\alpha_l\bE\left[A_l^2[t]+C_l^2[t]\right]$, $\epsilon$ is some positive constant satisfying 
$\boldsymbol{\lambda}+\epsilon\mb{1}\in\mc{R}$, and $\mb{1}$ is a vector of ones.
\end{proposition}
\begin{IEEEproof}
Consider the Lyapunov function
\begin{align}
W(\mb{Q}[t],\mb{T}[t])\triangleq\sum_{l=1}^{L}\alpha_lQ_l^2[t]+4\gamma C_{\max}\sum_{l=1}^{L}\beta_{l}T_l[t].
\end{align}
It is shown in Appendix \ref{app:proof:prop:throughput} that there exists a positive constant $\epsilon>0$ such that 
\begin{align}
\label{eqn:th:df:temp:main}
\Delta W\triangleq&\bE\left[W(\mb{Q}[t+1],\mb{T}[t+1])-W(\mb{Q}[t],\mb{T}[t])\middle|\mb{Q}[t],\mb{T}[t]\right]\nonumber\\
\leq&-2\epsilon\sum_{l=1}^{L}\alpha_lQ_l[t]+B(\boldsymbol{\alpha},\boldsymbol{\beta},\gamma).
\end{align}
Taking the expectation on the both sides of \eqref{eqn:th:df:temp:main} and summing over $t=0,1,...,K-1$, we have the desired result.
\end{IEEEproof}

Proposition~\ref{prop:throughput} establishes the throughput optimality
of the RSG Algorithm, thus $Q_l[t]$ and $T_l[t]$ will converge
in distribution to $\ol{Q}_l^*$ and $\ol{T}_l^*$,
which attain the steady-state distribution under our policy.
Proposition~\ref{prop:throughput} also gives an upper bound
for the expected total queue-length under the steady-state,
which increases linearly with the design parameter $\gamma$.
It will be revealed later that $\gamma$ controls the tradeoff
between the average total queue-length, and the service
regularity performance, especially in the heterogenous networks.  


Next, we will show that all moments of steady-state system variables, such as
queue-lengths and TLSL, are bounded under the RSG Algorithm, which enables us to
analyze the service regularity performance by using the
Lyapunov-type approach developed in \cite{erysri12}. In \cite{haj82}, the sufficient
condition for all moments of state variables of a Markov Chain to exist in steady state is given as finding a
Lyapunov function that satisfies: (1) it has a negative Lyapunov drift when the
system variable is large enough; (2) the absolute value of the
Lyapunov drift is bounded or has the exponential tail. Yet, the second condition
is hard to hold due to the unique evolution of TLSL counters, which have bounded
increment but unbounded decrement. We tackle this challenge by properly
partitioning the system space.
\begin{proposition}
\label{prop:moment}
For any arrival rate $\bs{\lambda}\in\text{Int}(\mc{R})$, all moments of
steady-state queue length and TLSL exist under the RSG Algorithm with any $\alpha_l>0$ and $\gamma>0$.
\end{proposition}
\begin{IEEEproof}
We show the boundedness of $\bE\left[e^{\eta\|\mb{Y}[t]\|_2}\right]$ for some $\eta>0$ by intelligently partitioning the system space, where $\mb{Y}[t]\triangleq\left(\sqrt{\boldsymbol{\alpha}}\mb{Q}[t],\sqrt{4\gamma C_{\max}\boldsymbol{\beta}\mb{T}[t]}\right)$, $\sqrt{\mb{x}}$ denotes the component-wise square root of the vector $\mb{x}$, and 
$\mb{xy}$ denotes the component-wise product of the vectors $\mb{x}$ and $\mb{y}$. Please 
see our technical report \cite{lieryilmazreg13} for details.
\end{IEEEproof}

Having established the throughput optimality and the moment existence of the system states of the RSG Algorithm, we are ready to analyze the service regularity performance, i.e., 
$\sum_{l=1}^{L}\beta_l\lambda_l\bE[\ol{T}_l]$. 

\section{Service Regularity Performance Analysis}
\label{sec:regular}
In this section, we study the service regularity performance of our proposed
RSG Algorithm analytically. We first establish a fundamental lower bound
on the service regularity for any feasible scheduling algorithm. Then, we derive
an upper bound on the service regularity under the RSG Algorithm. These investigations reveal that the service
regularity performance of the RSG Algorithm can be guaranteed
to remain within a factor of the lower bound, which is expressed as a function of
the system statistics and the design parameters, and can be as low as
$2$ in some special networks. We assume the parameter $\gamma>0$ throughout this section.

\subsection{Lower Bound Analysis}
\label{sec:lb}
In this subsection, we derive a lower bound based on a Lyapunov drift argument
inspired by the technique used in \cite{erysri12}. To study the lower bound of the service regularity by the Lyapunov drift argument, we consider a class of policies, called $\mc{P}$, that not only stabilize the system but also yield the bounded second moment of the steady-state TSLS\footnote{We conjecture that the second moment of the steady-state TSLS is bounded as long as the system is stable.} Note that our proposed algorithm, as well as the MWS algorithm,
falls into this class by Propositions~\ref{prop:throughput} and \ref{prop:moment}.

Let $\ol{T}_l^{(p)}$ and $\ol{S}_l^{(p)}$ be the steady-state TLSL and
scheduling variable for link $l$ under policy $p$, respectively. The following lemma
gives key identities for the first and second moment of the steady-state
TSLS, which are useful in deriving a lower bound on the service regularity.
\begin{lemma}
\label{lemma:identity}
For any policy $p\in\mc{P}$, we have
\begin{align}
\bE\left[\sum_{l\in\ol{\mb{H}}^{(p)}}\beta_l\lambda_l\ol{T}_l^{(p)}\right]=&\sum_{l=1}^{L}\beta_l\lambda_l-\bE\left[\sum_{l\in\ol{\mb{H}}^{(p)}}\beta_l\lambda_l\right],\label{eqn:T:first}\\
2\sum_{l=1}^{L}\beta_l\lambda_l\bE\left[\ol{T}_l^{(p)}\right]=&\sum_{l=1}^{L}\beta_l\lambda_l-\bE\left[\sum_{l\in\ol{\mb{H}}^{(p)}}\beta_l\lambda_l\right]\nonumber\\
&+\bE\left[\sum_{l\in\ol{\mb{H}}^{(p)}}\beta_l\lambda_l\left(\ol{T}_l^{(p)}\right)^2\right],\label{eqn:T:second}
\end{align}
where $\ol{\mb{H}}^{(p)}\triangleq\{l:\ol{C}_l\ol{S}_l^{(p)}>0\}$, and $\ol{\mb{C}}=(\ol{C}_l)_{l=1}^{L}$ has the same probability distribution as $\mb{C}[t]=(C_l[t])_{l=1}^{L}$.
\end{lemma}
\begin{IEEEproof}
See Appendix \ref{app:proof:lemma:identity} for the proof.
\end{IEEEproof}

We are ready to give a lower bound on the service regularity for any feasible policy $p\in\mc{P}$.
\begin{proposition}
\label{prop:lb}
For any policy $p\in\mc{P}$, we have
\begin{align*}
\sum_{l=1}^{L}\beta_l\lambda_l\bE\left[\ol{T}_l^{(p)}\right]\geq\frac{1}{2}\left(\frac{\sum_{l=1}^{L}\beta_l\lambda_l}{\max_{\mb{S}\in\mc{S}}\sum_{l\in\mb{S}}\beta_l\lambda_l}-1\right)\sum_{l=1}^{L}\beta_l\lambda_l.
\end{align*}
\end{proposition}
\begin{IEEEproof}
In the rest of proof, we will omit superscript $p$ for conciseness.
For any sample path, by Cauchy-Schwarz inequality, we have
\begin{align}
\left(\sum_{l\in\ol{\mb{H}}}\beta_l\lambda_l\ol{T}_l\right)^2=&\left(\sum_{l\in\ol{\mb{H}}}\sqrt{\beta_l\lambda_l}\cdot\sqrt{\beta_l\lambda_l}\ol{T}_l\right)^2\nonumber\\
\leq&\left(\sum_{l\in\ol{\mb{H}}}\beta_l\lambda_l\right)\sum_{l\in\ol{\mb{H}}}\beta_l\lambda_l\ol{T}_l^2,
\end{align}
where we recall that $\mb{\ol{H}}\triangleq\{l:\ol{C}_l\ol{S}_l>0\}$. This implies
\begin{align}
\sum_{l\in\ol{\mb{H}}}\beta_l\lambda_l\ol{T}_l^2\geq\frac{\left(\sum_{l\in\ol{\mb{H}}}\beta_l\lambda_l\ol{T}_l\right)^2}{\sum_{l\in\ol{\mb{H}}}\beta_l\lambda_l}.
\end{align}
Hence, we have
\begin{align}
\label{eqn:lb:key}
\bE\left[\sum_{l\in\ol{\mb{H}}}\beta_l\lambda_l\ol{T}_l^2\right]\geq&\bE\left[\frac{\left(\sum_{l\in\ol{\mb{H}}}\beta_l\lambda_l\ol{T}_l\right)^2}{\sum_{l\in\ol{\mb{H}}}\beta_l\lambda_l}\right]\nonumber\\
\stackrel{(a)}{\geq}&\frac{\left(\bE\left[\sum_{l\in\ol{\mb{H}}}\beta_l\lambda_l\ol{T}_l\right]\right)^2}{\bE\left[\sum_{l\in\ol{\mb{H}}}\beta_l\lambda_l\right]}\nonumber\\
\stackrel{(b)}{=}&\frac{\left(\sum_{l=1}^{L}\beta_l\lambda_l-\bE\left[\sum_{l\in\ol{\mb{H}}}\beta_l\lambda_l\right]\right)^2}{\bE\left[\sum_{l\in\ol{\mb{H}}}\beta_l\lambda_l\right]},
\end{align}
where the step $(a)$ uses the fact that $f(x,y)=\frac{x^2}{y}$ is convex
and Jensen's inequality for a multi-variable function;
step $(b)$ follows from \eqref{eqn:T:first}. By substituting \eqref{eqn:lb:key} into \eqref{eqn:T:second}, we have
\begin{align}
\label{eqn:lb:temp}
\sum_{l=1}^{L}\beta_l\lambda_l\bE\left[\ol{T}_l\right]\geq\frac{1}{2}\left(\frac{\sum_{l=1}^{L}\beta_l\lambda_l}{\bE\left[\sum_{l\in\ol{\mb{H}}}\beta_l\lambda_l\right]}-1\right)\sum_{l=1}^{L}\beta_l\lambda_l.
\end{align}
Note that
\begin{align}
\label{eqn:lb:up}
&\bE\left[\sum_{l\in\ol{\mb{H}}}\beta_l\lambda_l\right]=\bE\left[\sum_{l=1}^{L}\beta_l\lambda_l\mathds{1}_{\{\ol{C}_l\ol{S}_l>0\}}\right]\nonumber\\
=&\sum_{l=1}^{L}\beta_l\lambda_l\Pr\{\ol{C}_l\ol{S}_l>0\}\nonumber\\
\leq&\sum_{l=1}^{L}\beta_l\lambda_l\Pr\{\ol{S}_l=1\}
\leq\max_{\mb{S}\in\mc{S}}\sum_{l\in\mb{S}}\beta_l\lambda_l.
\end{align}
By substituting \eqref{eqn:lb:up} into \eqref{eqn:lb:temp}, we have the desired result.
\end{IEEEproof}

Consider a single-hop non-fading network, where only one link is scheduled in each time slot. Let $\beta_l=\beta$ and $\lambda_l=\lambda$ for each link $l$. Then, the lower bound becomes 
\begin{align}
\label{eqn:lb:rr}
\sum_{l=1}^{L}\bE\left[\ol{T}_l^{(p)}\right]\geq\frac{1}{2}L(L-1).
\end{align}

This lower bound can be achieved by the Round-Robin (RR) policy, which serves each link periodically. Thus, in the steady-state, the TSLS vector under the RR policy is a permutation of $\{0,1,2,...,L-1\}$ and thus $\sum_{l=1}^{L}\bE\left[\ol{T}_l^{(\text{RR})}\right]=\frac{1}{2}L(L-1).$

Yet, we would like to point out that the RR policy is not throughput-optimal.
Thus, for an arrival rate vector $\boldsymbol{\lambda}$ that cannot be supported by
the RR policy, we do not expect a throughput-optimal policy to approach
the above lower bound when serving it. However, for the arrival rate
vectors that can be supported by the RR policy, we shall see in our
numerical results that the performance of our policy can approach
 this lower bound when we increase the scaling parameter $\gamma$.

\subsection{Upper Bound Analysis}
\label{sec:ub}
In this subsection, we obtain an upper bound on the service regularity under the
RSG Algorithm. Let $\ol{Q}_l^{*}$, $\ol{S}_l^{*}$ and $\ol{T}_l^{*}$ be the steady-state queue-length, scheduling variable and TSLS for link $l$ under the RSG Algorithm, respectively.
\begin{proposition}
\label{prop:ub}
For the RSG Algorithm, we have
\begin{align}
\label{eqn:prop:ub}
\sum_{l=1}^{L}\beta_l\lambda_l\bE\left[\ol{T}_l^{*}\right]\leq& \frac{C_{\max}}{1+\epsilon}\left(\sum_{l=1}^{L}\beta_l-\bE\left[\sum_{l\in\ol{\mb{H}}^{*}}\beta_l\right]\right)\nonumber\\
&+\frac{1}{2\gamma(1+\epsilon)}\sum_{l=1}^{L}\alpha_l\bE\left[\ol{A}_l^2+\ol{C}_l^2\right],
\end{align}
where $\epsilon>0$ satisfies $\bs{\lambda}(1+\epsilon)\in\mc{R}$, 
$\ol{\mb{H}}^{*}\triangleq\{l:\ol{C}_l\ol{S}_l^{*}>0\}$, and $\ol{\mb{A}}=(\ol{A}_l)_{l=1}^{L}$ has the same distribution as $\mb{A}[t]=(A_l[t])_{l=1}^{L}$.
\end{proposition}
\begin{IEEEproof}
See Appendix \ref{app:proof:prop:ub} for the details.
\end{IEEEproof}

Note that the second term of the right hand side of (\ref{eqn:prop:ub}) captures various random effects in the network: the burstiness of the arrival processes and the channel variations. Under our policy these effects diminish as the scaling factor $\gamma$ goes to infinity. Hence, together with Proposition~\ref{prop:throughput}, Proposition~\ref{prop:ub} reveals a tradeoff: when increasing $\gamma$, the upper bound on the total queue-length increases linearly with $\gamma$, but the upper bound for the service regularity decreases.

Consider the single-hop non-fading network as in Section \ref{sec:lb}. Let $\beta_l=\beta$ and 
$\lambda_l=\lambda=\frac{1}{L(1+\epsilon)}$ for each link $l$. Then, as $\gamma$ goes to infinity, \eqref{eqn:prop:ub} becomes 
\begin{align}
\sum_{l=1}^{L}\bE\left[\ol{T}_l^{*}\right]\leq L(L-1),
\end{align}
which is always within twice the value of the lower bound \eqref{eqn:lb:rr}. In the more general case, the upper bound converges to a constant that is determined by the system statistics and design parameters as $\gamma$ goes to infinity. Moreover, we shall see in the numerical results presented in Section~\ref{sec:simRegulartiy} that as $\gamma$ increases, the service regularity performance under the RSG Algorithm actually converges to the lower bound \eqref{eqn:lb:rr} in the single-hop non-fading network with the symmetric parameters.

\begin{figure*}[htbp]
\centering \subfloat[Single-hop non-fading network]{
\label{fig:sim_2by2switch_sym}
\includegraphics[scale=0.42]{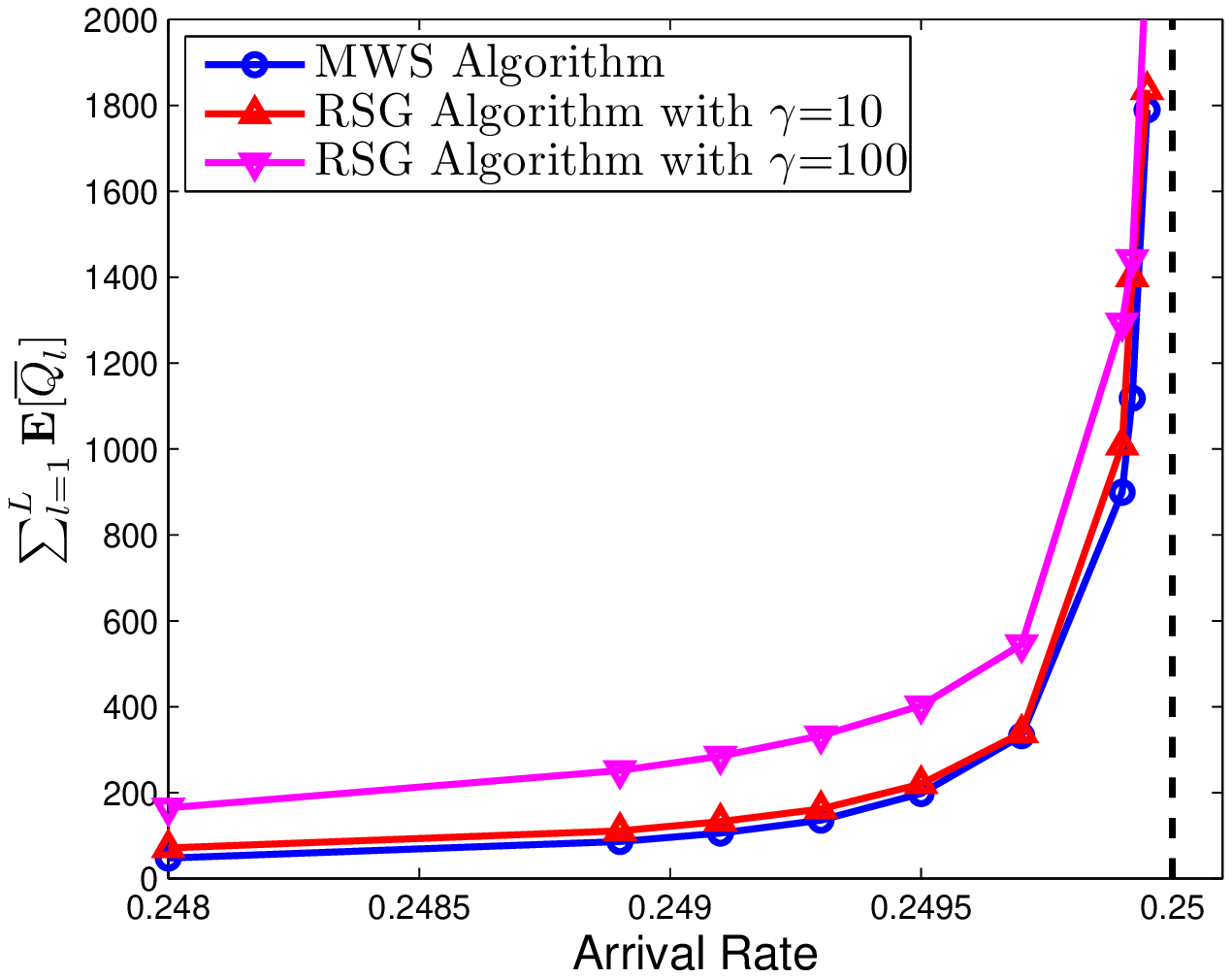}
} \hspace{-0.3in}\subfloat[Single-hop fading network]{
\label{fig:sim_2by2switch_asym}
\includegraphics[scale=0.42]{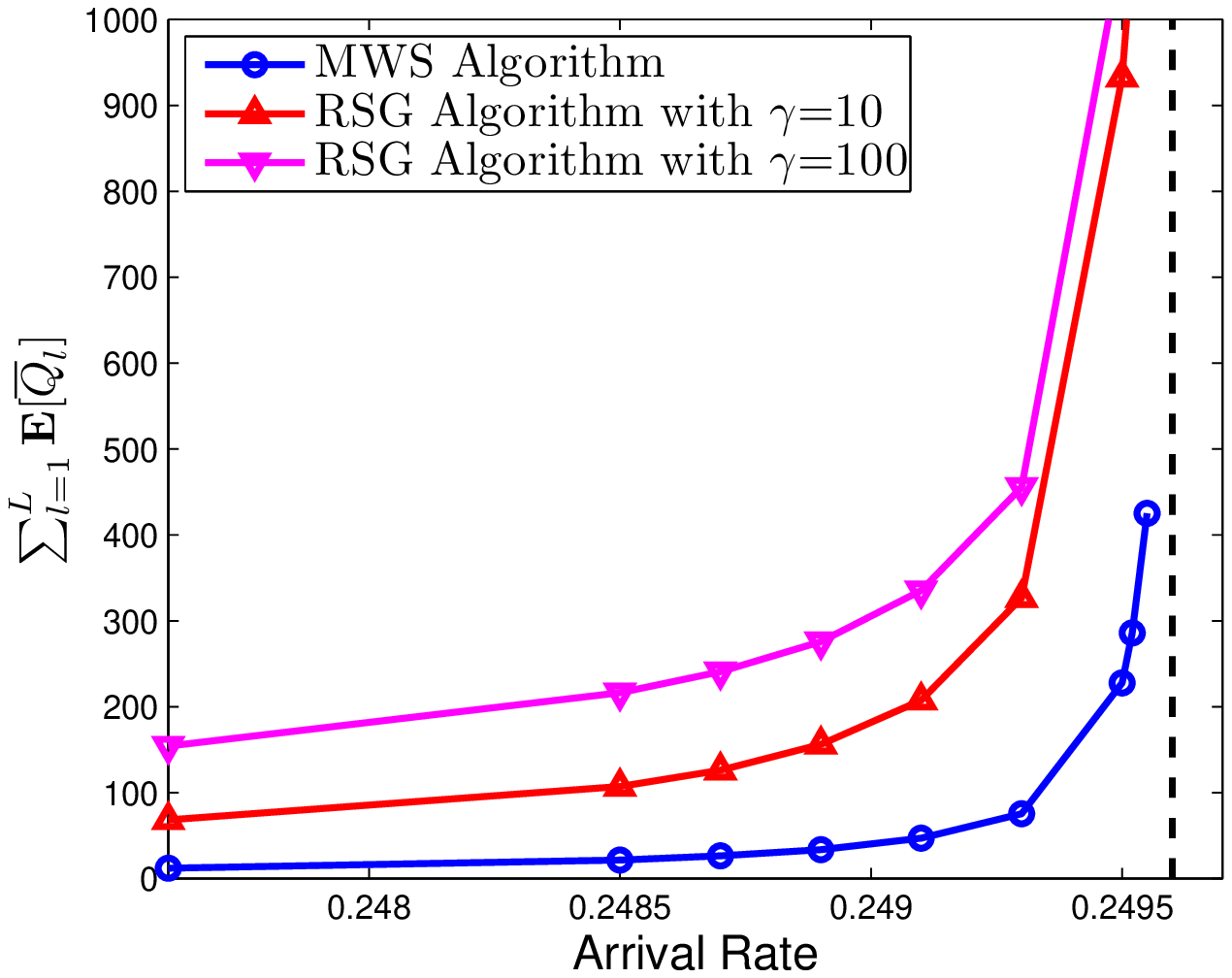}
}\hspace{-0.3in} \subfloat[$3\times 3$ switch]{ \label{fig:sim_3by3switch}
\includegraphics[scale=0.42]{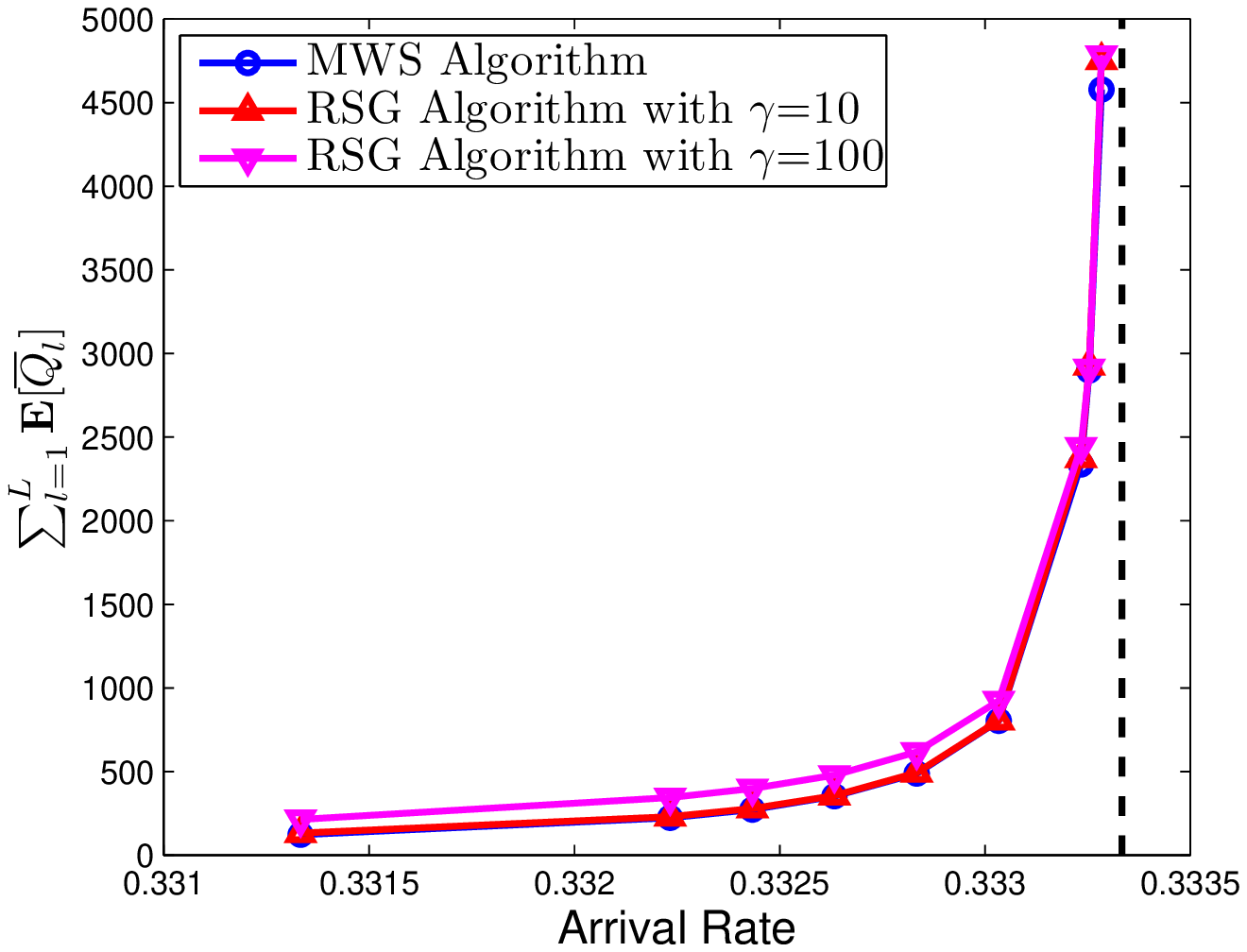}
} \caption{The throughput performance of the RSG Algorithm}
\label{fig:throughput}
\end{figure*}

\begin{figure*}[htbp]
\centering \subfloat[Single-hop non-fading network]{
\label{fig:sim_2by2switch_sym}
\includegraphics[scale=0.31]{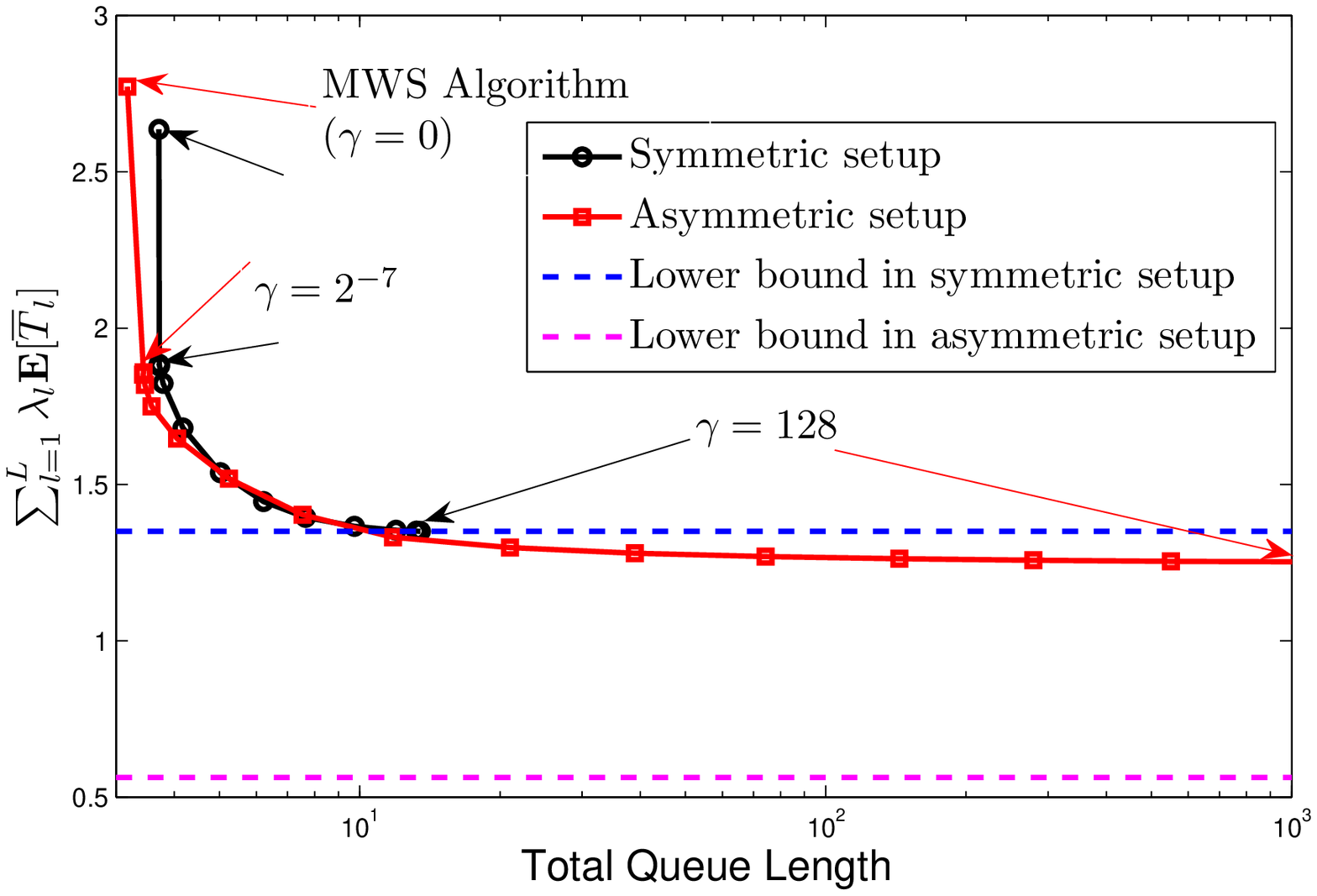}
} \hspace{-0.3in}\subfloat[Single-hop fading network]{
\label{fig:sim_2by2switch_asym}
\includegraphics[scale=0.31]{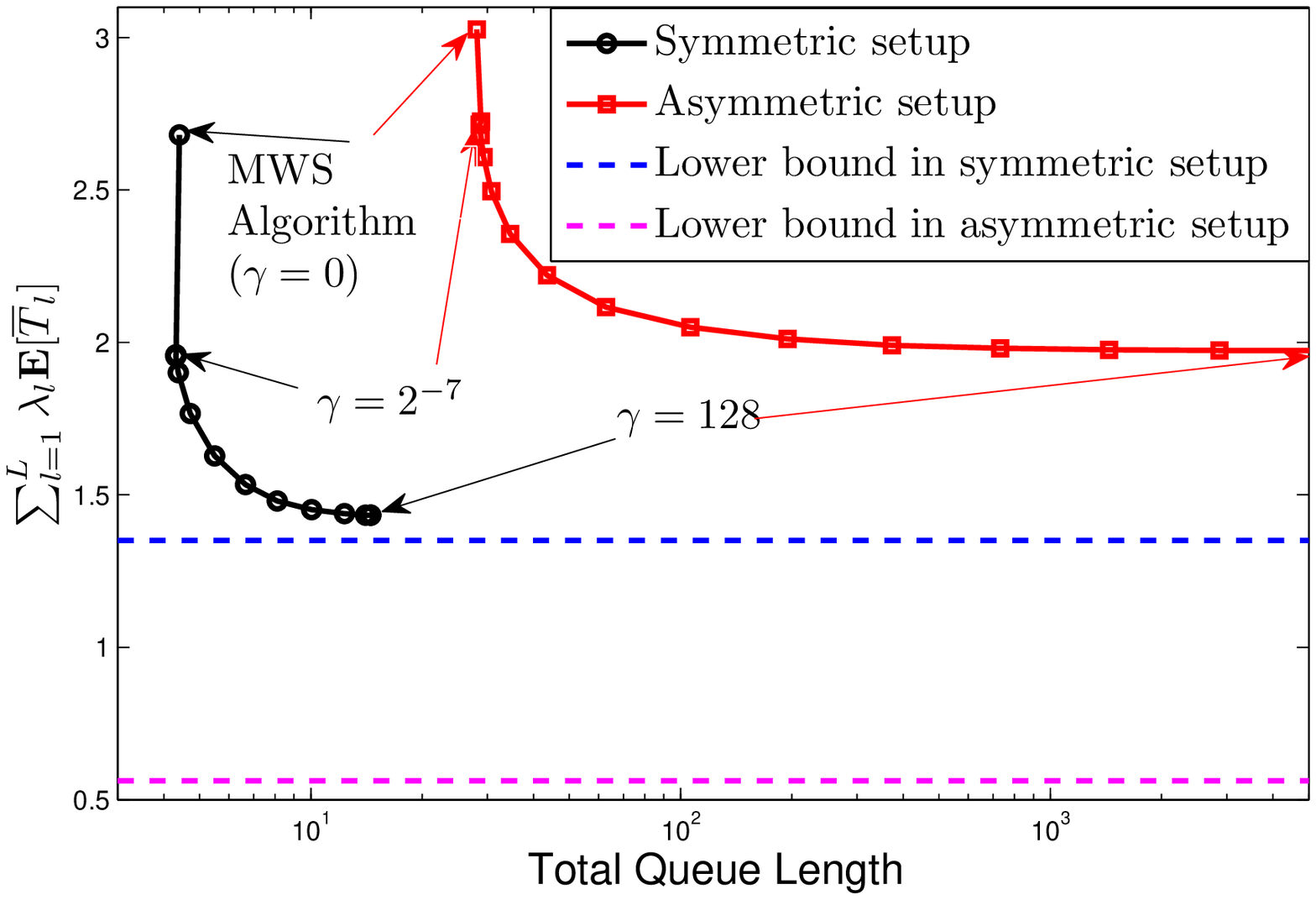}
}\hspace{-0.3in} \subfloat[$3\times 3$ switch]{ \label{fig:sim_3by3switch}
\includegraphics[scale=0.31]{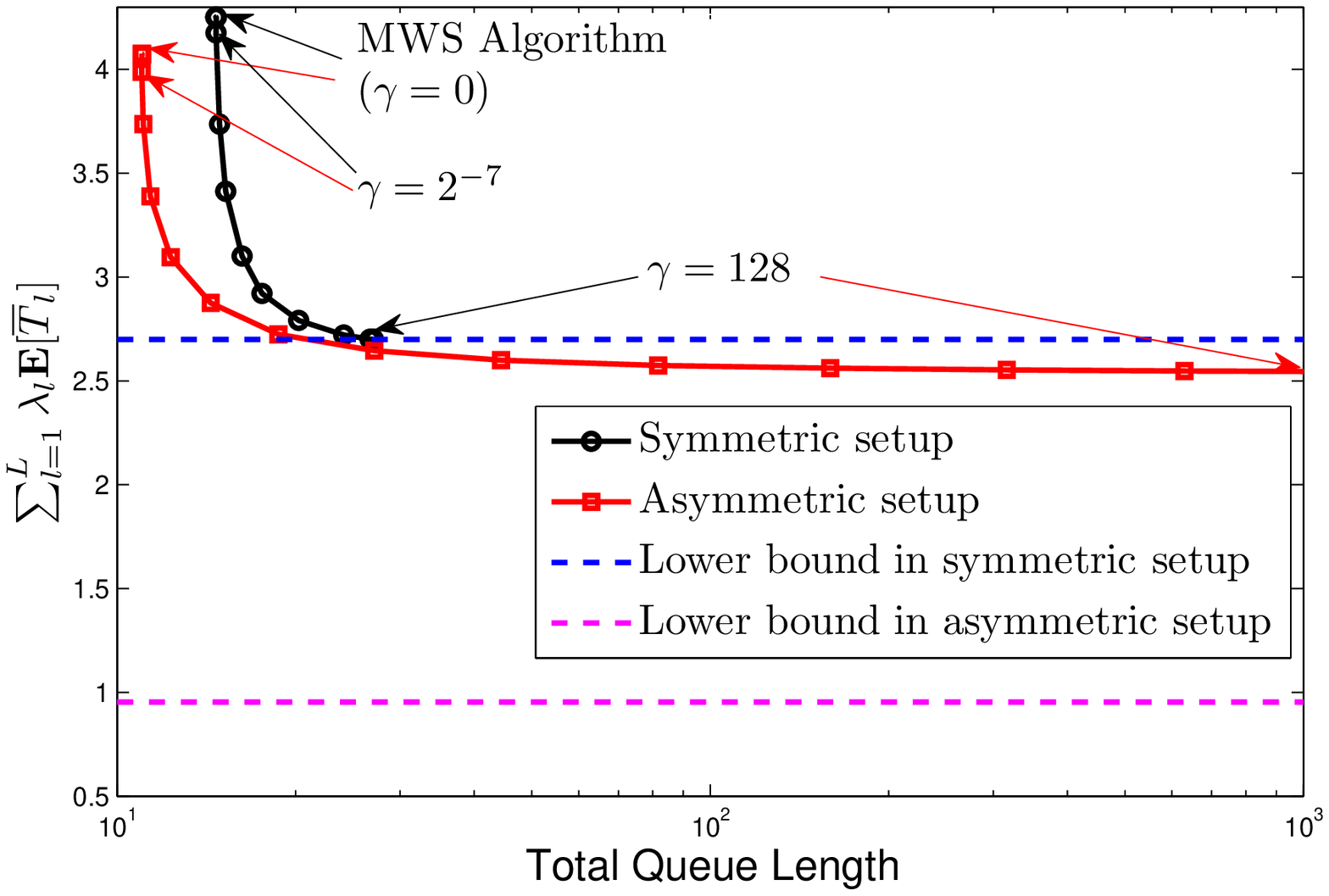}
} \caption{Trade-off between mean queue length and the service regularity}
\label{fig:regularity}
\end{figure*}

\section{Numerical Results}\label{sec:simulation}

In this section, we provide simulation results for our proposed RSG Algorithm and compare its performance to the MWS Algorithm and bounds. In addition to investigating the throughput (cf. Section~\ref{sec:simThroughput}) and service regularity (cf. Section~\ref{sec:simRegulartiy}) performances of our policy in both single-hop network with $L=4$ links and $3\times3$ switch, we also look at the behavior of the RSG Algorithm as well as the potential benefit of the service regularity (cf. Section~\ref{sec:simBenefit}). In the first two simulations, we assume Bernoulli arrivals to each link and $\alpha_l=\beta_l=1$ for each link $l$.

\subsection{Throughput Performance}\label{sec:simThroughput}

In this subsection, we illustrate the throughput performance of the RSG Algorithm in three different network setups with symmetric arrivals: (i) single-hop non-fading network, (ii) single-hop network with symmetric ON-OFF fading channels with probability $q=0.8$ that the channel is available, and (iii) $3\times3$ switch. The achievable rate regions for these three networks, respectively, are
\begin{align*} 
\Lambda_1\triangleq&\left\{\boldsymbol{\lambda}=(\lambda_l)_{l=1}^{4}:\lambda_1=\lambda_2=...=\lambda_4<\frac{1}{4}\right\}, \nonumber\\
\Lambda_2\triangleq&\left\{\boldsymbol{\lambda}=(\lambda_l)_{l=1}^{4}:\lambda_1=\lambda_2=...=\lambda_4<\frac{1-(1-q)^{4}}{4}\right\},\nonumber\\
\Lambda_3\triangleq&\left\{\boldsymbol{\lambda}=(\lambda_l)_{l=1}^{9}:\lambda_1=\lambda_2=...=\lambda_9<\frac{1}{3}\right\}.
\end{align*} 

In Fig.~\ref{fig:throughput}, we compare the total mean queue-length under the MWS Algorithm, as well as the RSG Algorithm with different $\gamma$ values. It can be observed in Fig.~\ref{fig:throughput} that the RSG Algorithm can stabilize the system in the above network setups.
It also can be seen that the total mean queue-length of the RSG Algorithm increases with the parameter $\gamma$. This is expected since as $\gamma$ increases, it becomes more likely for the RSG Algorithm to choose a queue with less packet to serve, potentially wasting some service while improving the service regularity, as we shall see next.


\subsection{Service Regularity Performance}\label{sec:simRegulartiy}

In this subsection, we investigate the service regularity performance of our RSG Algorithm, as well as illustrate the tradeoff between the total mean queue-length and the service regularity. We present our results in three different networks: single-hop non-fading network, single-hop fading network and $3\times3$ switch. In both single-hop nonfading and fading networks, we consider the symmetric setup with the arrival rate vector $\boldsymbol{\lambda}\triangleq[0.225, 0.225, 0.225, 0.225]$, and the asymmetric setup with the arrival rate vector $\boldsymbol{\lambda}\triangleq[0.4, 0.3, 0.15, 0.05]$. For a single-hop ON-OFF fading network, the probability vectors that the channels are available are
$\mb{q}=[0.8, 0.8, 0.8, 0.8]$ in symmetric setup and $\mb{q}=[0.6, 0.5, 0.4, 0.3]$ in asymmetric setup. For a $3\times3$ switch, we consider the symmetric setup with the arrival rate vector $\boldsymbol{\lambda}\triangleq[0.3, 0.3, 0.3; 0.3, 0.3, 0.3; 0.3, 0.3, 0.3]$ and the asymmetric setup with the arrival rate vector $\boldsymbol{\lambda}\triangleq[0.5, 0.3, 0.1; 0.2, 0.4, 0.3; 0.1, 0.2, 0.5]$.
In all simulations, we choose the scaling parameter $\gamma$ to be the powers of 2, ranging from $2^{-7}$ to $2^7$.

Fig.~\ref{fig:regularity} shows the relationship between the total mean queue-length and the service regularity in different network setups. The tradeoff between the service regularity and the total mean queue-length can be clearly seen: as $\gamma$ increases, the service regularity improves while the total mean queue-length also increases. It can be observed that the simulated values converge to the fundamental lower bound in non-fading networks with symmetric setup (Figs~\ref{fig:regularity}(a) and (c)), while they stay away from the lower bound in asymmetric setups. This motivates us to refine the lower bound analysis in asymmetric setups, which is left for future investigation. Here, it is worth mentioning that even with very small $\gamma$ values (e.g., $2^{-6}$), our RSG Algorithm significantly improves the service regularity, while introducing negligible increase in the total mean queue-length.

\begin{figure*}[htb!]
\centering 
\subfloat[Mean unused service]{
\label{fig:RSG:MWS:unuse}
\hspace{-0.2in}\includegraphics[scale=0.33]{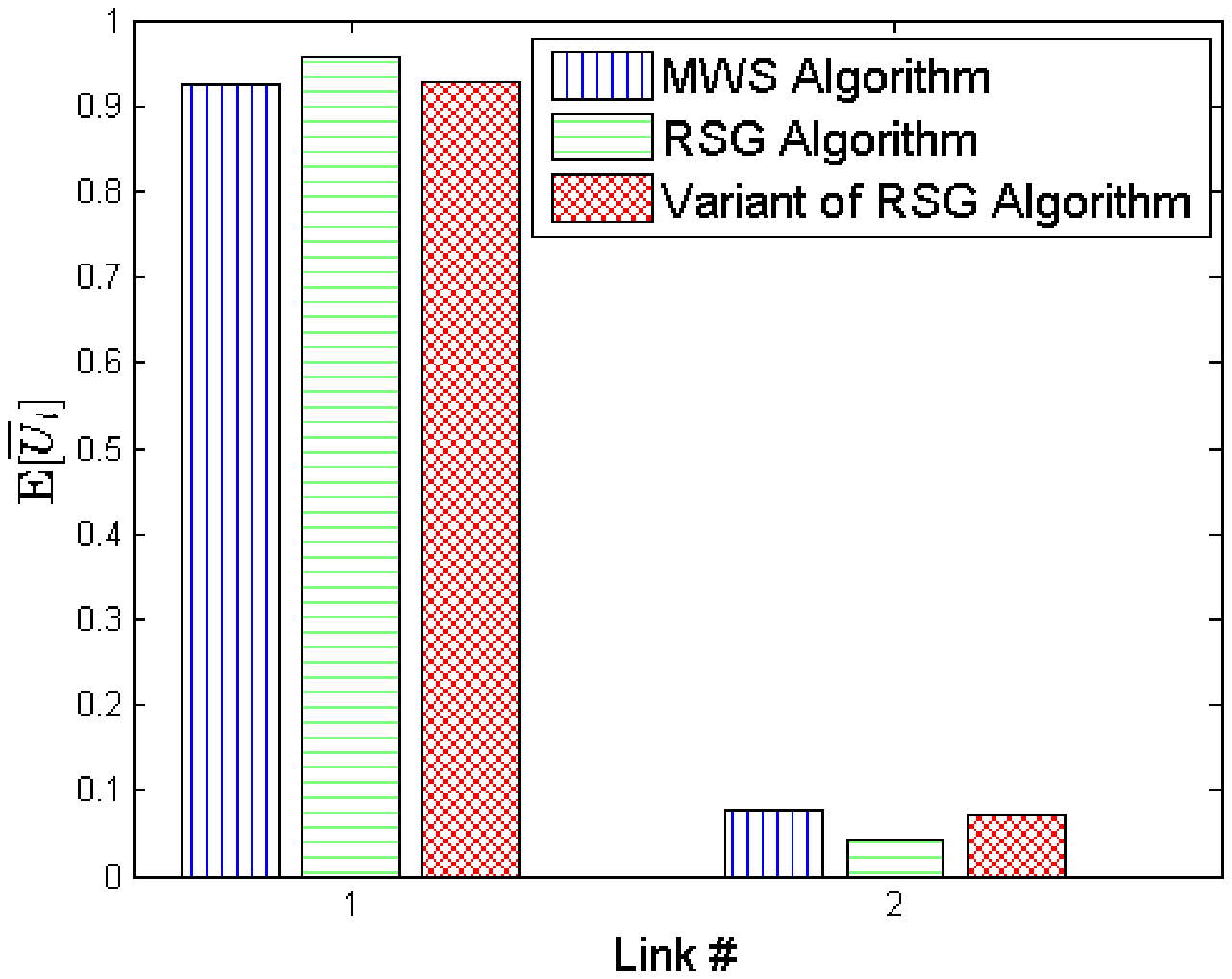}
} \subfloat[Service regularity performance]{
\label{fig:RSG:MWS:regularity}
\hspace{-0.2in}\includegraphics[scale=0.28]{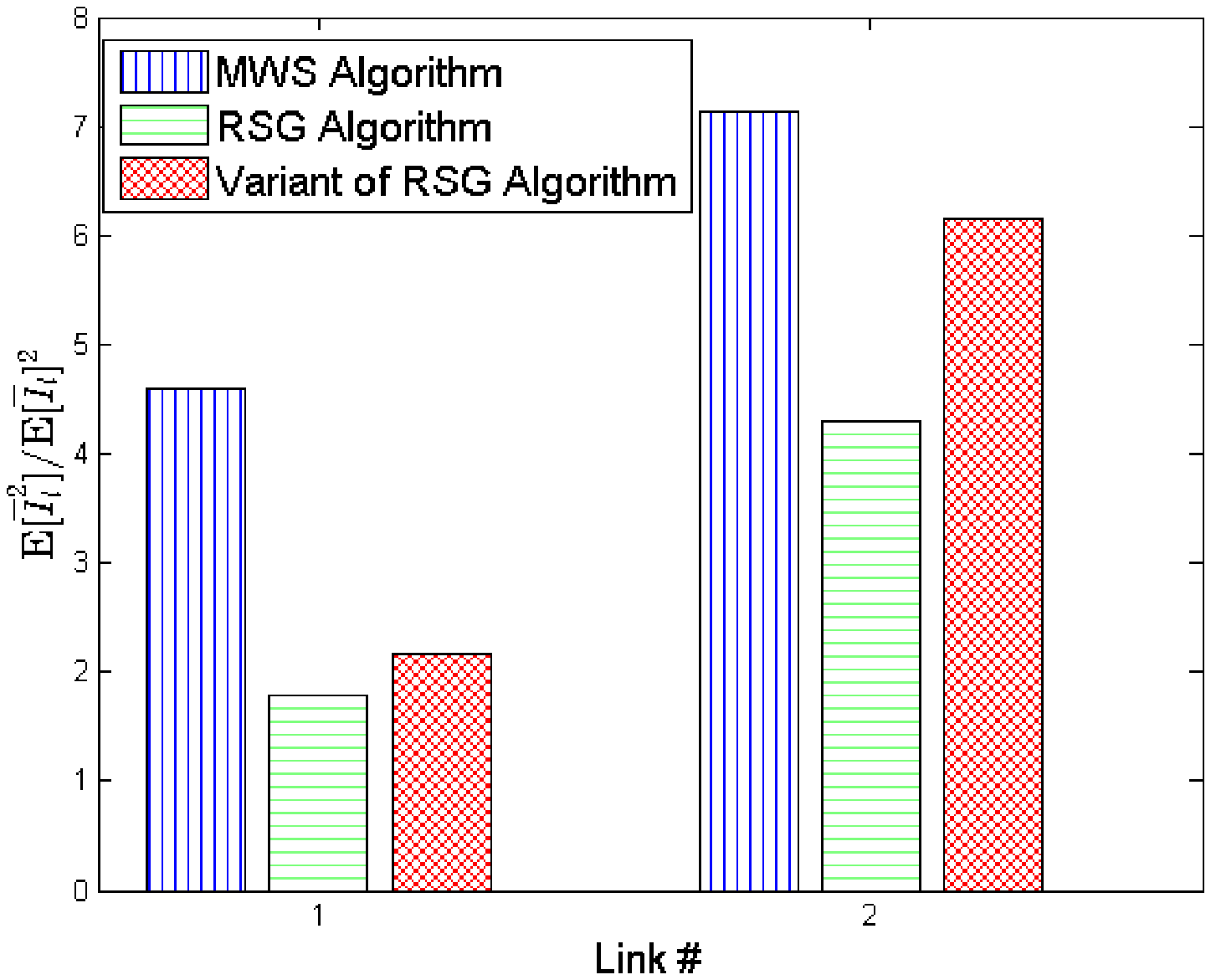}
} 
\subfloat[Mean queue length]{
\label{fig:RSG:MWS:meanQ}
\hspace{-0.2in}\includegraphics[scale=0.33]{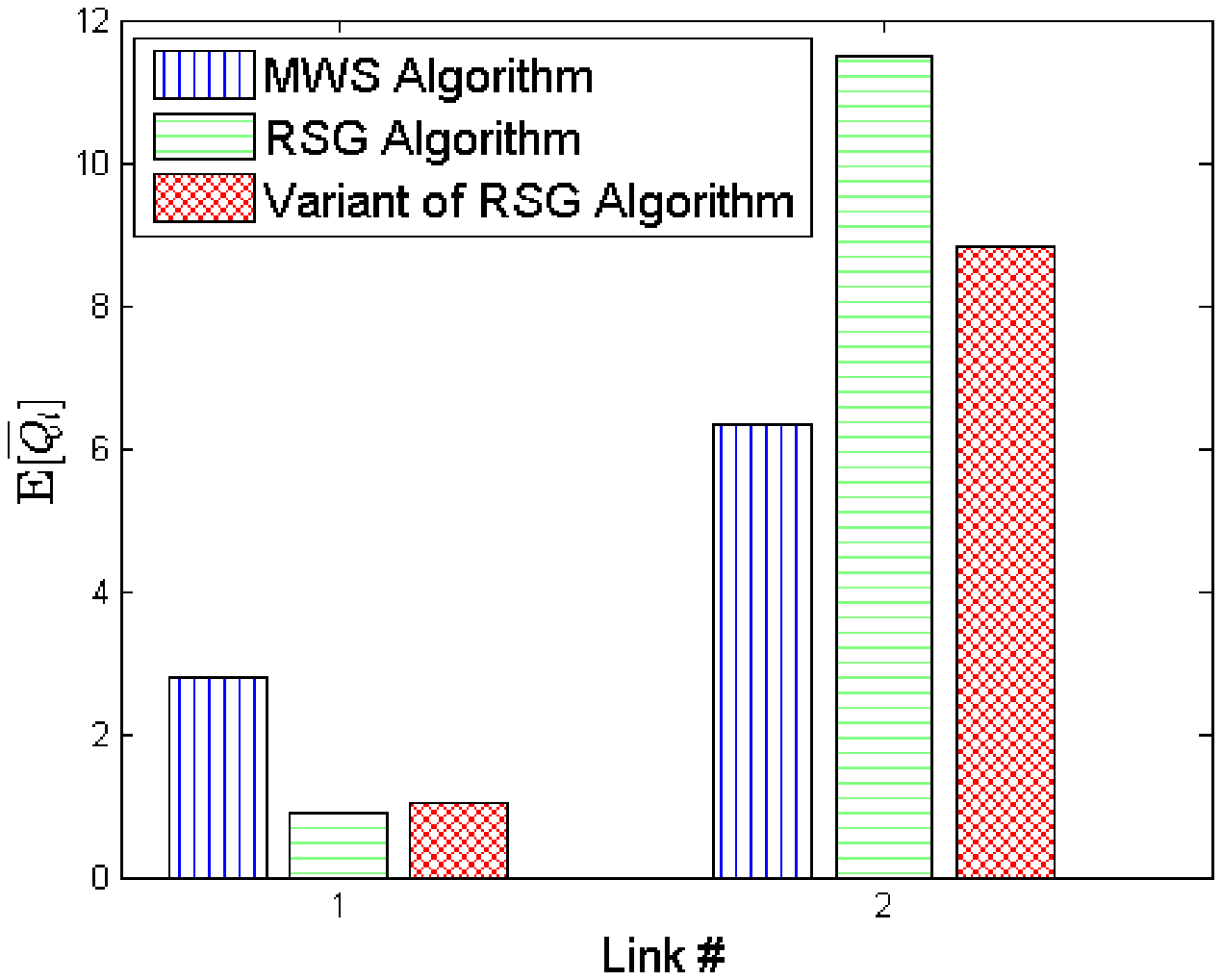}
}
\subfloat[Standard deviation of queue length]{
\label{fig:RSG:MWS:varQ}
\hspace{-0.2in}\includegraphics[scale=0.33]{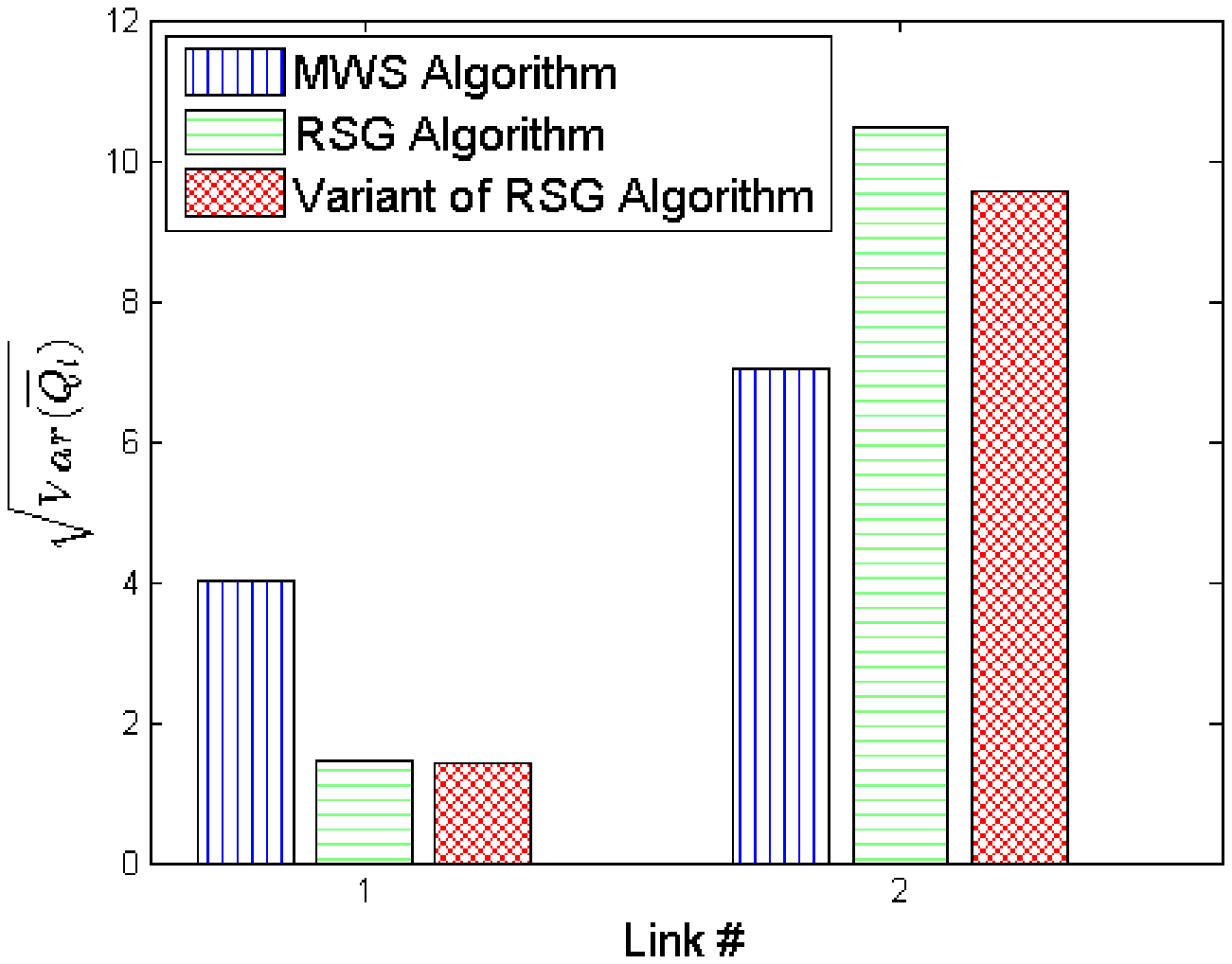}
} \caption{Performance comparison between the RSG Algorithm and the MWS Algorithm}
\end{figure*}

\subsection{Benefit of the Service Regularity}
\label{sec:simBenefit}

In this subsection, we study various performance metrics (such as mean unused service, service regularity, mean queue-length and variance of queue-length) among links to illustrate the behavior of the RSG Algorithm as well as the benefit of the service regularity. To that end, we consider a single-hop non-fading network with two links. Each link can serve $4$ packets in each time slot if scheduled. There is always $1$ packet arriving at the first link in each time slot, while the number of packets arriving at the second link is either $2K$ with probability $1/K$ or $0$, where $K$ is a natural number. 
We compare the performance between the MWS Algorithm, the RSG Algorithm and the variant of the RSG Algorithm whose TSLS counter increases only when the link does not receive service and the link queue-length is non-zero. In both RSG Algorithm and its variant, we set $\beta_1=1, \beta_2=0, \gamma=10$, i.e., we assume that the first link prefers the regular service while the second link does not have such a requirement. In the following simulations, we set $K=5$. 

From Fig. \ref{fig:RSG:MWS:unuse}, we observe that compared to the MWS Algorithm, under the RSG Algorithm, the mean unused service in the first link slightly increases, while in the second it slightly decreases. This is expected since the TSLS counter increases even when the queue-length is non-zero. Yet, the total amount of mean unused service under the RSG Algorithm remains the same as that under the MWS Algorithm. For the variant of the RSG Algorithm, the mean unused service for each individual link almost does not change. 

From Fig. \ref{fig:RSG:MWS:regularity}, we can see that both the RSG Algorithm and its variant improve the service regularity compared to the MWS Algorithm. Also, the RSG Algorithm yields the better service regularity performance than its variant. This is because the TSLS counter under the variant of the RSG Algorithm is not as aggressive as that under the original RSG Algorithm. As can be seen in Fig. \ref{fig:RSG:MWS:meanQ} and Fig. \ref{fig:RSG:MWS:varQ}, providing more regular service is extremely beneficial for the link with constant arrivals since it leads to the smaller mean and variance of delay that each packet experiences in that link.  

\begin{figure}[htb!]
\begin{center}
\vspace{-0.3in}
\includegraphics[scale=0.4]{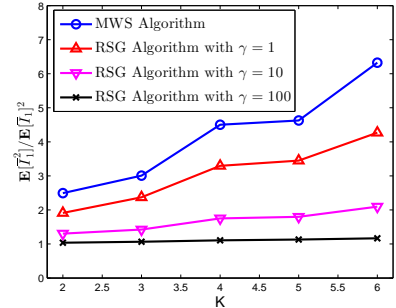}
\caption{The impact of the bursty arrivals on the service regularity of the constant flow}\vspace{-0.3in}
\label{fig:regular:bursty}
\end{center}
\end{figure}

Next, we would like to reveal the relationship between the service regularity of the first link with constant arrivals and the burstiness of arrivals at the second link that is reflected by the parameter $K$. The larger the $K$, the more bursty the arrivals at the second link. Fig. \ref{fig:regular:bursty} shows the impact of the bursty arrivals on the service regularity of the link with the constant arrival under both MWS and RSG Algorithms. We can observe from Fig. \ref{fig:regular:bursty} that the service regularity of the first link under the MWS Algorithm degrades much faster than that under the RSG Algorithm as the the burstiness of the second link increases. Also, as $\gamma$ increases, under the RSG Algorithm, the service regularity of the first link improves significantly, and it is almost independent of the burstiness of the second link when $\gamma=100$.

\section{Conclusion}\label{sec:conclusion}

In this work, we investigated the problem of designing a scheduling policy that is both throughput-optimal and possesses favorable service regularity characteristics. We introduced a new parameter of time-since-last-service, and proposed a novel scheduling policy that combines this parameter with the queue-lengths in its weight.
After establishing the throughput optimality of our policy, we showed that it also has provable service regularity performance. In particular, the service regularity of our policy can be guaranteed to remain within a factor distance of a fundamental lower bound for any feasible scheduling policy. We explicitly expressed this factor as a function of the system statistics and the design parameters. We performed extensive numerical studies to illustrate the significant gains achieved by our policy over the traditional queue-length-based policies. Our results show the significance of utilizing the time-since-last-service in improving the service regularity performance of throughput-optimal policies.



\appendices
\section{Proof of Lemma \ref{lmm:TandI}}
\label{app:proofLmmTandI}
Without loss of generality, assume that link $l$ is served at time $0$. For any positive integer $M$, there exists an $m$ such that
\begin{eqnarray*}
I_l[1]+\dots+I_l[m] &\leq& M,\\
I_l[1]+\dots+I_l[m]+I_l[m+1] &>& M.
\end{eqnarray*}
We can write
\begin{eqnarray}
\frac{1}{M}\sum_{t=1}^M T_l[t]
&=&\frac{1}{M} \Bigg(\sum_{t=1}^{I_l[1]}T_l[t] + \sum_{t=I_l[1]+1}^ {I_l[1]+I_l[2]} T_l[t] + \dots\nonumber\\
&& + \sum_{t=I_l[1]+\dots+I_l[m]+1}^ M T_l[t] \Bigg). \label{eqn:proofLmmTandI10}
\end{eqnarray}

We observe the following fact: assume link $l$ receives its $(m-1)^{th}$ and $m^{th}$ service at
time slot $t_1$ and $t_2$, respectively, where $t_2>t_1$. Then, by definition,
$I_l[m]=t_2-t_1$, $T_l[t_1+1] = 0$ and $T_l[t_2]=t_2-t_1-1=I_l[m]-1$.
Using this fact, we know the $k^{th}$ summation on the right hand
 side of (\ref{eqn:proofLmmTandI10}) gives
 $\frac{1}{2}(I_l[k](I_l[k]-1))$, except for the last one.
 Thus we have:
\begin{eqnarray}
\frac{1}{M}\sum_{t=1}^M T_l[t] &\geq& \frac{1}{M} \sum_{k=1}^m \frac{I_l[k](I_l[k]-1)}{2}, \label{eqn:proofLmmTandI20}\\
\frac{1}{M}\sum_{t=1}^M T_l[t] &\leq& \frac{1}{M} \sum_{k=1}^{m+1} \frac{I_l[k](I_l[k]-1)}{2}. \label{eqn:proofLmmTandI30}
\end{eqnarray}

By the definition of $m$ and the fact that $\bE[\overline{I}_l]<\infty$ when
each link is served with strictly positive probability,
we know that $m\rightarrow\infty$ when $M\rightarrow\infty$.
Since the policy considered in this paper is Markovian, we have
\begin{eqnarray*}
\lim_{M\rightarrow\infty} \frac{1}{M} \sum_{t=1}^M T_l[t]= \bE[\overline{T}_l],\\
\lim_{m\rightarrow\infty} \frac{1}{m} \sum_{k=1}^m I_l[k]= \bE[\overline{I}_l],\\
\lim_{m\rightarrow\infty} \frac{1}{m} \sum_{k=1}^m I_l^2[k]= \bE[\overline{I}_l^2].
\end{eqnarray*}
Note that
\begin{eqnarray*}
\frac{1}{m} \sum_{k=1}^m I_l[k] \leq \frac{M}{m} \leq \frac{m+1}{m}\frac{1}{m+1}\sum_{k=1}^{m+1} I_l[k],
\end{eqnarray*}
which implies $\lim_{M\rightarrow\infty} \frac{M}{m}= \bE[\overline{I}_l].$

By taking limit on (\ref{eqn:proofLmmTandI20}) and (\ref{eqn:proofLmmTandI30}) as $M\rightarrow\infty$, we have
\begin{eqnarray*}
&& \lim_{M\rightarrow\infty} \frac{m}{M} \frac{1}{m} \sum_{k=1}^m \frac{I_l[k](I_l[k]-1)}{2}\\
&\leq& \lim_{M\rightarrow\infty} \frac{1}{M} \sum_{t=1}^M T_l[t]\\
&\leq& \lim_{M\rightarrow\infty} \frac{m+1}{M} \frac{1}{m+1} \sum_{k=1}^{m+1} \frac{I_l[k](I_l[k]-1)}{2},
\end{eqnarray*}
and thus we have the desired result.

\section{Proof of inequality \eqref{eqn:th:df:temp:main}}
\label{app:proof:prop:throughput}
\begin{align}
\label{eqn:th:df:temp}
\Delta W\triangleq&\bE\left[W(\mb{Q}[t+1],\mb{T}[t+1])-W(\mb{Q}[t],\mb{T}[t])\middle|\mb{Q}[t],\mb{T}[t]\right]\nonumber\\
=&\bE\Bigg[\sum_{l=1}^{L}\alpha_lQ_l^2[t+1]+4\gamma C_{\max}\sum_{l=1}^{L}\beta_lT_l[t+1]\nonumber\\
&-\sum_{l=1}^{L}\alpha_lQ_l^2[t]-4\gamma C_{\max}\sum_{l=1}^{L}\beta_lT_l[t]\Bigg|\mb{Q}[t],\mb{T}[t]\Bigg]\nonumber\\
\leq&\sum_{l=1}^{L}\alpha_l\bE\left[(Q_l[t]+A_l[t]-C_l[t]S_l^{*}[t])^2-Q_l^2[t]\middle|\mb{Q}[t],\mb{T}[t]\right]\nonumber\\
&+4\gamma C_{\max}\bE\left[\sum_{l=1}^{L}\beta_lT_l[t+1]-\sum_{l=1}^{L}\beta_lT_l[t]\middle|\mb{Q}[t],\mb{T}[t]\right],
\end{align}
where the last step follows from the evolution of each queue, and
$(\max\{x,0\})^2\leq x^2$.

Let $\mb{H}^{*}\triangleq\{l: S^{*}_l[t]C_l[t]>0\}$. According to the
definition of the TSLS counter, we have
\begin{align}
&\sum_{l=1}^{L}\beta_lT_l[t+1]\nonumber\\
=&\sum_{l\notin\mb{H}^{*}}\beta_l\left(T_l[t]+1\right)\nonumber\\
=&\sum_{l=1}^{L}\beta_lT_l[t]-\sum_{l\in\mb{H}^{*}}\beta_lT_l[t]+\sum_{l=1}^{L}\beta_l-\sum_{l\in\mb{H}^{*}}\beta_l\label{eqn:th:T:identity}\\
\leq&\sum_{l=1}^{L}\beta_lT_l[t]-\sum_{l\in\mb{H}^{*}}\beta_lT_l[t]+\sum_{l=1}^{L}\beta_l. \label{eqn:th:Tev}
\end{align}
By substituting inequality \eqref{eqn:th:Tev} into \eqref{eqn:th:df:temp}, we have
\begin{align}
\label{eqn:th:df:temp:sim}
&\Delta W\leq\nonumber\\
&\sum_{l=1}^{L}\alpha_l\bE\left[(Q_l[t]+A_l[t]-C_l[t]S_l^{*}[t])^2-Q_l^2[t]\middle|\mb{Q}[t],\mb{T}[t]\right]\nonumber\\
&+4\gamma C_{\max}\bE\left[\sum_{l=1}^{L}\beta_l-\sum_{l\in\mb{H}^{*}}\beta_lT_l[t]\middle|\mb{Q}[t],\mb{T}[t]\right]\nonumber\\
\leq&\sum_{l=1}^{L}\alpha_l\bE\left[2Q_l[t](A_l[t]-C_l[t]S_l^{*}[t])\middle|\mb{Q}[t],\mb{T}[t]\right]\nonumber\\
&+\sum_{l=1}^{L}\alpha_l\bE\left[(A_l[t]-C_l[t]S_l^{*}[t])^2\middle|\mb{Q}[t],\mb{T}[t]\right]\nonumber\\
&+4\gamma C_{\max}\sum_{l=1}^{L}\beta_l-4\gamma C_{\max}\bE\left[\sum_{l\in\mb{H}^{*}}\beta_lT_l[t]\middle|\mb{Q}[t],\mb{T}[t]\right]\nonumber\\
\leq&2\sum_{l=1}^{L}\alpha_l\lambda_{l}Q_l[t]-2\bE\left[\sum_{l=1}^{L}\alpha_lQ_l[t]C_l[t]S_l^{*}[t]\middle|\mb{Q}[t],\mb{T}[t]\right]\nonumber\\
&-4\gamma C_{\max}\bE\left[\sum_{l\in\mb{H}^{*}}\beta_lT_l[t]\middle|\mb{Q}[t],\mb{T}[t]\right]+B(\boldsymbol{\alpha},\boldsymbol{\beta},\gamma),
\end{align}
where $B(\boldsymbol{\alpha},\boldsymbol{\beta},\gamma)$ is defined in Proposition \ref{prop:throughput}.

Let $\displaystyle\mb{S}^{(\text{MWS})}[t]\in\argmax_{\mb{S}\in\mc{S}}\sum_{l=1}^{L}\alpha_lQ_l[t]C_l[t]S_l$. Then, by the definition of the RSG Algorithm, we have
\begin{align*}
&\sum_{l=1}^{L}(\alpha_lQ_l[t]+\gamma\beta_lT_l[t])C_l[t]S_l^{*}[t]\nonumber\\
\geq&\sum_{l=1}^{L}(\alpha_lQ_l[t]+\gamma\beta_lT_l[t])C_l[t]S_l^{(\text{MWS})}[t]\nonumber\\
\geq&\sum_{l=1}^{L}\alpha_lQ_l[t]C_l[t]S_l^{(\text{MWS})}[t],
\end{align*}
which implies
\begin{align}
\label{eqn:th:df:mw}
&\sum_{l=1}^{L}\alpha_lQ_l[t]C_l[t]S_l^{*}[t]\nonumber\\
\geq&\sum_{l=1}^{L}\alpha_lQ_l[t]C_l[t]S_l^{(\text{MWS})}[t]-\gamma\sum_{l=1}^{L}\beta_lT_l[t]C_l[t]S_l^{*}[t].
\end{align}
By substituting \eqref{eqn:th:df:mw} into \eqref{eqn:th:df:temp:sim}, we have
\begin{align}
\label{eqn:th:df:afT}
&\Delta W\leq2\sum_{l=1}^{L}\alpha_l\lambda_{l}Q_l[t]-2\bE\left[\sum_{l=1}^{L}\alpha_lQ_l[t]C_l[t]S_l^{(\text{MWS})}[t]\middle|\mb{Q},\mb{T}\right]\nonumber\\
&+2\gamma\bE\left[\sum_{l=1}^{L}\beta_lT_l[t]C_l[t]S_l^{*}[t]\middle|\mb{Q}[t],\mb{T}[t]\right]\nonumber\\
&-4\gamma C_{\max}\bE\left[\sum_{l\in\mb{H}^{*}}\beta_lT_l[t]\middle|\mb{Q}[t],\mb{T}[t]\right]+B(\boldsymbol{\alpha},\boldsymbol{\beta},\gamma).
\end{align}
Given $\mb{Q}[t]$ and $\mb{T}[t]$, we have 
\begin{align}
\label{eqn:th:df:T}
C_{\max}\bE\left[\sum_{l\in\mb{H}^{*}}\beta_lT_l[t]\middle|\mb{Q},\mb{T}\right]\geq\bE\left[\sum_{l=1}^{L}\beta_lT_l[t]C_l[t]S_l^{*}[t]\middle|\mb{Q},\mb{T}\right], 
\end{align}
where we recall that $\mb{H}^{*}=\{l:S_l^{*}[t]C_l[t]>0\}$. 
By substituting \eqref{eqn:th:df:T} into \eqref{eqn:th:df:afT}, we have
\begin{align}
\label{eqn:th:df:temp:last}
&\Delta W\leq2\sum_{l=1}^{L}\alpha_l\lambda_{l}Q_l[t]-2\bE\left[\sum_{l=1}^{L}\alpha_lQ_l[t]C_l[t]S_l^{(\text{MWS})}[t]\middle|\mb{Q},\mb{T}\right]\nonumber\\
&-2\gamma\bE\left[\sum_{l=1}^{L}\beta_lT_l[t]C_l[t]S_l^{*}[t]\middle|\mb{Q}[t],\mb{T}[t]\right]+B(\boldsymbol{\alpha},\boldsymbol{\beta},\gamma).
\end{align}
Note that the capacity region $\mc{R}$ (see \cite{tas97}) is also equivalent to a set of arrival rate
vectors $\bs{\lambda}$ such that there exist non-negative numbers $\theta(\mb{c};\mb{s})$ satisfying
\begin{align}
\lambda_l\leq\sum_{\mb{c}}\Pr\{\mb{C}[t]=\mb{c}\}\sum_{\mb{s}\in\mc{S}}\theta(\mb{c};\mb{s})c_{l}s_{l},\forall l,
\end{align}
where $\mb{s}=(s_l)_{l=1}^{L}$ and $\sum_{\mb{s}\in\mc{S}}\theta(\mb{c};\mb{s})=1,\forall \mb{c}$. For any $\bs{\lambda}\in\text{Int}(\mc{R})$, there exists an $\epsilon>0$ such that
\begin{align}
\lambda_l\leq\sum_{\mb{c}}\Pr\{\mb{C}[t]=\mb{c}\}\sum_{\mb{s}\in\mc{S}}\theta(\mb{c};\mb{s})c_{l}s_{l}-\epsilon,\forall l.
\end{align}
Hence, we have
\begin{align}
\label{eqn:th:df:cp}
&\sum_{l=1}^{L}\alpha_l\lambda_{l}Q_l[t]+\epsilon\sum_{l=1}^{L}\alpha_lQ_l[t]\nonumber\\
\leq&\sum_{\mb{c}}\Pr\{\mb{C}[t]=\mb{c}\}\sum_{\mb{s}\in\mc{S}}\theta(\mb{c};\mb{s})\sum_{l=1}^{L}\alpha_lQ_l[t]c_{l}s_l\nonumber\\
\stackrel{(a)}{\leq}&\sum_{\mb{c}}\Pr\{\mb{C}[t]=\mb{c}\}\sum_{\mb{s}\in\mc{S}}\theta(\mb{c};\mb{s})\sum_{l=1}^{L}\alpha_lQ_l[t]c_{l}S_l^{(\text{MWS})}[t]\nonumber\\
=&\bE\left[\sum_{l=1}^{L}\alpha_lQ_l[t]C_l[t]S_l^{(\text{MWS})}[t]\middle|\mb{Q}[t],\mb{T}[t]\right].
\end{align}
where the step (a) follows from the definition of $\mb{S}^{(\text{MWS})}$.
By substituting \eqref{eqn:th:df:cp} into \eqref{eqn:th:df:temp:last}, we have
\begin{align}
\Delta W\leq&-2\epsilon\sum_{l=1}^{L}\alpha_lQ_l[t]+B(\boldsymbol{\alpha},\boldsymbol{\beta},\gamma)\nonumber\\
-&2\gamma\bE\left[\sum_{l=1}^{L}\beta_lT_l[t]C_l[t]S_l^{*}[t]\middle|\mb{Q}[t],\mb{T}[t]\right]\label{eqn:mt:ineq}\\
\leq&-2\epsilon\sum_{l=1}^{L}\alpha_lQ_l[t]+B(\boldsymbol{\alpha},\boldsymbol{\beta},\gamma).  \label{eqn:th:final}
\end{align}

\section{Proof of Lemma \ref{lemma:identity}}
\label{app:proof:lemma:identity}
In the rest of proof, we will omit the superscript $p$ for brevity.

\textbf{Proof of identity \eqref{eqn:T:first}}:
\begin{align}
&\sum_{l=1}^{L}\beta_l\lambda_{l}T_l[t+1]=\sum_{l\notin\mb{H}}\beta_l\lambda_l\left(T_l[t]+1\right)\nonumber\\
=&\sum_{l=1}^{L}\beta_l\lambda_{l}T_l[t]-\sum_{l\in\mb{H}}\beta_l\lambda_{l}T_l[t]+\sum_{l=1}^{L}\beta_l\lambda_l-\sum_{l\in\mb{H}}\beta_l\lambda_l,
\end{align}
where $\mb{H}\triangleq\{l:S_l[t]C_l[t]>0\}$. 
Taking expectation on both sides with respect to the steady state distribution of
$(\mb{Q},\mb{T})$ and rearranging terms, we have the desired result.

\textbf{Proof of identity \eqref{eqn:T:second}}:
\begin{align}
&\sum_{l=1}^{L}\beta_l\lambda_{l}T_l^2[t+1]=\sum_{l\notin\mb{H}}\beta_l\lambda_l\left(T_l[t]+1\right)^2\nonumber\\
=&\sum_{l\notin\mb{H}}\beta_l\lambda_lT_l^2[t]+2\sum_{l\notin\mb{H}}\beta_l\lambda_lT_l[t]+\sum_{l\notin\mb{H}}\beta_l\lambda_l\nonumber\\
=&\sum_{l=1}^{L}\beta_l\lambda_lT_l^2[t]-\sum_{l\in\mb{H}}\beta_l\lambda_lT_l^2[t]+2\sum_{l=1}^{L}\beta_l\lambda_lT_l[t]\nonumber\\
&-2\sum_{l\in\mb{H}}\beta_l\lambda_lT_l[t]+\sum_{l=1}^{L}\beta_l\lambda_l-\sum_{l\in\mb{H}}\beta_l\lambda_l.
\end{align}
Taking expectation on both sides with respect to the steady state distribution of
$(\mb{Q},\mb{T})$ and rearranging terms, we have
\begin{align}
2\sum_{l=1}^{L}\beta_l\lambda_l\bE\left[\ol{T}_l\right]=&2\bE\left[\sum_{l\in\ol{\mb{H}}}\beta_l\lambda_l\ol{T}_l\right]+\bE\left[\sum_{l\in\ol{\mb{H}}}\beta_l\lambda_l\ol{T}_l^2\right]\nonumber\\
&-\left(\sum_{l=1}^{L}\beta_l\lambda_l-\bE\left[\sum_{l\in\ol{\mb{H}}}\beta_l\lambda_l\right]\right).
\end{align}
Using Identity \eqref{eqn:T:first}, we have the desired result.

\section{Proof of Proposition \ref{prop:ub}}
\label{app:proof:prop:ub}
Consider the quadratic Lyapunov function
$W_{\mb{Q}}(\mb{Q},\mb{T})\triangleq\frac{1}{2}\sum_{l=1}^{L}\alpha_lQ_l^2$.
We have
\begin{align}
&\Delta W_{\mb{Q}}(\mb{Q},\mb{T})\nonumber\\
=&\bE\left[W_{\mb{Q}}(\mb{Q}[t+1],\mb{T}[t+1])-W_{\mb{Q}}(\mb{Q}[t],\mb{T}[t])\middle|\mb{Q}[t],\mb{T}[t]\right]\nonumber\\
=&\bE\left[\frac{1}{2}\sum_{l=1}^{L}\alpha_lQ_l^2[t+1]-\frac{1}{2}\sum_{l=1}^{L}\alpha_lQ_l^2[t]\middle|\mb{Q}[t],\mb{T}[t]\right]\nonumber\\
\leq&\frac{1}{2}\sum_{l=1}^{L}\bE\left[\alpha_l\left(Q_l[t]+A_l[t]-C_l[t]S_l^{*}[t]\right)^2-\alpha_lQ_l^2[t]\middle|\mb{Q}[t],\mb{T}[t]\right]\nonumber\\
\leq&\sum_{l=1}^{L}\alpha_l\bE\left[Q_l[t]\left(A_l[t]-C_l[t]S_l^{*}[t]\right)\middle|\mb{Q}[t],\mb{T}[t]\right]\nonumber\\
&+\frac{1}{2}\sum_{l=1}^{L}\alpha_l\bE\left[A_l^2[t]+C_l^2[t]\right].
\end{align}
Taking expectation on both sides with respect to the steady state distribution
of $(\mb{Q},\mb{T})$, and using the fact that
$\bE[\Delta W_{\mb{Q}}(\ol{\mb{Q}},\ol{\mb{T}})]=0$ followed from
$\bE[\ol{Q}_l^2]<\infty$ for all $l\in\mc{L}$, we have
\begin{align}
0\leq&\sum_{l=1}^{L}\alpha_l\lambda_l\bE\left[\ol{Q}_l^{*}\right]-\sum_{l=1}^{L}\alpha_l\bE\left[\ol{Q}_l^{*}\ol{C}_l\ol{S}_l^{*}\right]\nonumber\\
&+\frac{1}{2}\sum_{l=1}^{L}\alpha_l\bE\left[\ol{A}_l^2+\ol{C}_l^2\right],
\end{align}
which implies
\begin{align*}
\sum_{l=1}^{L}\alpha_l\bE\left[\ol{Q}_l^{*}\ol{C}_l\ol{S}_l^{*}\right]\leq\sum_{l=1}^{L}\alpha_l\lambda_l\bE\left[\ol{Q}_l^{*}\right]+\frac{1}{2}\sum_{l=1}^{L}\alpha_l\bE\left[\ol{A}_l^2+\ol{C}_l^2\right].
\end{align*}
Hence, we have
\begin{align}
\label{eqn:up:main}
&\sum_{l=1}^{L}\bE\left[\left(\alpha_l\ol{Q}_l^{*}+\gamma\beta_l\ol{T}_l^{*}\right)\ol{C}_l\ol{S}_l^{*}\right]\nonumber\\
\leq&\sum_{l=1}^{L}\alpha_l\lambda_l\bE\left[\ol{Q}_l^{*}\right]
+\gamma\sum_{l=1}^{L}\beta_l\bE\left[\ol{T}_l^{*}\ol{S}_l^{*}\ol{C}_l\right]+\frac{1}{2}\sum_{l=1}^{L}\alpha_l\bE\left[\ol{A}_l^2+\ol{C}_l^2\right].
\end{align}
Recall that given $\mb{Q}[t]=\mb{Q}$, $\mb{T}[t]=\mb{T}$ and the channel
state $\mb{C}[t]$, we have
\begin{align}
&\sum_{l=1}^{L}\left(\alpha_lQ_l[t]+\gamma\beta_l T_l[t]\right)C_l[t]S_l^{*}[t]\nonumber\\
=&\max_{\mb{S}\in\mc{S}}\sum_{l=1}^{L}\left(\alpha_lQ_l[t]+\gamma\beta_lT_l[t]\right)C_l[t]S_l.
\end{align}
According to the definition of the capacity region $\mc{R}$, we can show 
\begin{align}
\label{eqn:RSG:fading:eq}
&\sum_{l=1}^{L}\bE\left[\left(\alpha_lQ_l[t]+\gamma\beta_lT_l[t]\right)C_l[t]S_l^{*}[t]\middle|\mb{Q}[t]=\mb{Q},\mb{T}[t]=\mb{T}\right]\nonumber\\
=&\max_{\mb{r}\in\mc{R}}\sum_{l=1}^{L}\left(\alpha_lQ_l[t]+\gamma\beta_lT_l[t]\right)r_l,
\end{align}
The proof is available in Appendix \ref{app:proof:eqn:RSG}

Since $\bs{\lambda}\in\text{Int}(\mc{R})$, there exists an $\epsilon>0$ such that $\bs{\lambda}(1+\epsilon)\in\mc{R}$. Hence, we have
\begin{align*}
&\sum_{l=1}^{L}\bE\left[\left(\alpha_lQ_l[t]+\gamma\beta_lT_l[t]\right)C_l[t]S_l^{*}[t]\middle|\mb{Q}[t]=\mb{Q},\mb{T}[t]=\mb{T}\right]\nonumber\\
\geq&\sum_{l=1}^{L}\lambda_l(1+\epsilon)\bE\left[\alpha_lQ_l[t]+\gamma\beta_lT_l[t]\middle|\mb{Q}[t]=\mb{Q},\mb{T}[t]=\mb{T}\right].
\end{align*}
Taking expectation on both sides with respect to the steady state distribution
of $(\mb{Q},\mb{T})$, we have
\begin{align*}
&\sum_{l=1}^{L}\bE\left[\left(\alpha_l\ol{Q}_l^{*}+\gamma\beta_l\ol{T_l}^{*}\right)\ol{C}_l\ol{S}_l^{*}\right]\nonumber\\
\geq&\sum_{l=1}^{L}\lambda_l(1+\epsilon)\bE\left[\alpha_l\ol{Q}_l^{*}+\gamma\beta_l\ol{T}_l^{*}\right].
\end{align*}
By substituting above inequality into \eqref{eqn:up:main} and canceling the
common term in both sides, we have
\begin{align*}
&\sum_{l=1}^{L}\beta_l\lambda_l\bE\left[\ol{T}_l^{*}\right]\nonumber\\
\leq&\frac{1}{1+\epsilon}\sum_{l=1}^{L}\beta_l\bE\left[\ol{T}_l^{*}\ol{S}_l^{*}\ol{C}_l\right]+\frac{1}{2\gamma(1+\epsilon)}\sum_{l=1}^{L}\alpha_l\bE\left[\ol{A}_l^2+\ol{C}_l^2\right]\nonumber\\
\leq&\frac{C_{\max}}{1+\epsilon}\bE\left[\sum_{l\in\ol{\mb{H}}^{*}}\beta_l\ol{T}_l^{*}\right]+\frac{1}{2\gamma(1+\epsilon)}\sum_{l=1}^{L}\alpha_l\bE\left[\ol{A}_l^2+\ol{C}_l^2\right]\nonumber\\
=&\frac{C_{\max}}{1+\epsilon}\left(\sum_{l=1}^{L}\beta_l-\bE\left[\sum_{l\in\ol{\mb{H}}^{*}}\beta_l\right]\right)\nonumber\\
&+\frac{1}{2\gamma(1+\epsilon)}\sum_{l=1}^{L}\alpha_l\bE\left[\ol{A}_l^2+\ol{C}_l^2\right],
\end{align*}
where the last step uses identity \eqref{eqn:T:first}.

\section{Proof of Equation \eqref{eqn:RSG:fading:eq}}
\label{app:proof:eqn:RSG}
We will use the following fact in linear programming. 

\begin{align}
\label{eqn:linear:prog}
\max_{\mb{x}\in\mc{A}}\sum_{l=1}^{L}a_lx_l=\max_{\mb{x}\in\text{CH}\{\mc{A}\}}\sum_{l=1}^{L}a_lx_l,
\end{align}
where $\mb{x}=(x_l)_{l=1}^{L}$ is a $L-$dimensional vector, $\mc{A}$ is a set of $L-$ dimensional vectors, $\text{CH}\{\mc{A}\}$ is a convex hull of the set $\mc{A}$ and $a_l,\forall l=1,2,...,L$, are real numbers. 

Given $\mb{Q}[t]$, $\mb{T}[t]$ and $\mb{C}[t]$, we have 
\begin{align}
\label{eqn:RSG:anotherform}
&\sum_{l=1}^{L}(\alpha_lQ_l[t]+\gamma\beta_lT_l[t])C_l[t]S_l^{*}[t]\nonumber\\
=&\max_{\mb{S}\in\mc{S}}\sum_{l=1}^{L}(\alpha_lQ_l[t]+\gamma\beta_lT_l[t])C_l[t]S_l\nonumber\\
=&\max_{\mb{v}=(v_l)_{l=1}^{L}\in\mc{S}^{(\mb{C}[t])}}\sum_{l=1}^{L}(\alpha_lQ_l[t]+\gamma\beta_lT_l[t])v_l,
\end{align}
where we recall that $\mc{S}^{(\mb{c})}\triangleq\{\mb{S}\mb{c}:\mb{S}\in\mc{S}\}$, and $\mb{a}\mb{b}\triangleq(a_lb_l)_{l=1}^{L}$ denotes the component-wise product of two vectors $\mb{a}$ and $\mb{b}$. 

Next, we will show that 
\begin{align}
&\sum_{l=1}^{L}\bE[(\alpha_lQ_l[t]+\gamma\beta_lT_l[t])C_l[t]S_l^{*}[t]|\mb{Q}[t]=\mb{Q},\mb{T}[t]=\mb{T}]\nonumber\\
=&\max_{\mb{r}=(r_l)_{l=1}^{L}\in\mc{R}}\sum_{l=1}^{L}(\alpha_lQ_l[t]+\gamma\beta_lT_l[t])r_l.
\end{align}

On one hand, 
\begin{align}
\label{eqn:pf:one}
&\sum_{l=1}^{L}\bE[(\alpha_lQ_l[t]+\gamma\beta_lT_l[t])C_l[t]S_l^{*}[t]|\mb{Q}[t]=\mb{Q},\mb{T}[t]=\mb{T}]\nonumber\\
\stackrel{(a)}{=}&\bE\left[\max_{\mb{v}\in\mc{S}^{(\mb{C}[t])}}\sum_{l=1}^{L}(\alpha_lQ_l[t]+\gamma\beta_lT_l[t])v_l\middle|\mb{Q}[t]=\mb{Q},\mb{T}[t]=\mb{T}\right]\nonumber\\
=&\sum_{\mb{c}}\Pr\{\mb{C}[t]=\mb{c}\}\max_{\mb{v}\in\mc{S}^{(\mb{c})}}\sum_{l=1}^{L}(\alpha_lQ_l[t]+\gamma\beta_lT_l[t])v_l\nonumber\\
\stackrel{(b)}{=}&\sum_{\mb{c}}\Pr\{\mb{C}[t]=\mb{c}\}\max_{\mb{v}\in\text{CH}\{\mc{S}^{(\mb{c})}\}}\sum_{l=1}^{L}(\alpha_lQ_l[t]+\gamma\beta_lT_l[t])v_l\nonumber\\
\stackrel{(c)}{=}&\sum_{\mb{c}}\Pr\{\mb{C}[t]=\mb{c}\}\sum_{l=1}^{L}(\alpha_lQ_l[t]+\gamma\beta_lT_l[t])v_l^{*(\mb{c})}\nonumber\\
=&\sum_{l=1}^{L}(\alpha_lQ_l[t]+\gamma\beta_lT_l[t])\sum_{\mb{c}}\Pr\{\mb{C}[t]=\mb{c}\}v_l^{*(\mb{c})}\nonumber\\
\stackrel{(d)}{\leq}&\max_{\mb{r}=(r_l)_{l=1}^{L}\in\mc{R}}\sum_{l=1}^{L}(\alpha_lQ_l[t]+\gamma\beta_lT_l[t])r_l,
\end{align}
where the steps (a) and (b) follow from equation \eqref{eqn:RSG:anotherform} and equation \eqref{eqn:linear:prog}, respectively; step (c) is true for 
\begin{align*}
\mb{v}^{*(\mb{c})}=(v_l^{*(\mb{c})})_{l=1}^{L}\in\argmax_{\mb{v}\in\text{CH}\{\mc{S}^{(\mb{c})}\}}\sum_{l=1}^{L}(\alpha_lQ_l[t]+\gamma\beta_lT_l[t])v_l;
\end{align*}
and step (d) follows from the fact that $\mb{v}^{*(\mb{c})}\in\text{CH}\{\mc{S}^{(\mb{c})}\}$ and $\sum_{\mb{c}}\Pr\{\mb{C}[t]=\mb{c}\}\mb{v}^{*(\mb{c})}\in\mc{R}$.

On the other hand, 
\begin{align}
\label{eqn:pf:second}
&\max_{\mb{r}=(r_l)_{l=1}^{L}\in\mc{R}}\sum_{l=1}^{L}(\alpha_lQ_l[t]+\gamma\beta_lT_l[t])r_l\nonumber\\
\stackrel{(a)}{=}&\sum_{l=1}^{L}(\alpha_lQ_l[t]+\gamma\beta_lT_l[t])r_l^{*}\nonumber\\
\stackrel{(b)}{=}&\sum_{l=1}^{L}(\alpha_lQ_l[t]+\gamma\beta_lT_l[t])\sum_{\mb{c}}\Pr\{\mb{C}[t]=\mb{c}\}v_l^{(\mb{c})}\nonumber\\
=&\sum_{\mb{c}}\Pr\{\mb{C}[t]=\mb{c}\}\sum_{l=1}^{L}(\alpha_lQ_l[t]+\gamma\beta_lT_l[t])v_l^{(\mb{c})}\nonumber\\
\leq&\sum_{\mb{c}}\Pr\{\mb{C}[t]=\mb{c}\}\max_{\mb{v}\in\text{CH}\{\mc{S}^{(\mb{c})}\}}\sum_{l=1}^{L}(\alpha_lQ_l[t]+\gamma\beta_lT_l[t])v_l\nonumber\\
\stackrel{(c)}{=}&\sum_{\mb{c}}\Pr\{\mb{C}[t]=\mb{c}\}\max_{\mb{v}\in\mc{S}^{(\mb{c})}}\sum_{l=1}^{L}(\alpha_lQ_l[t]+\gamma\beta_lT_l[t])v_l\nonumber\\
=&\bE\left[\max_{\mb{v}\in\mc{S}^{(\mb{C}[t])}}\sum_{l=1}^{L}(\alpha_lQ_l[t]+\gamma\beta_lT_l[t])v_l\middle|\mb{Q}[t]=\mb{Q},\mb{T}[t]=\mb{T}\right]\nonumber\\
\stackrel{(d)}{=}&\sum_{l=1}^{L}\bE[(\alpha_lQ_l[t]+\gamma\beta_lT_l[t])C_l[t]S_l^{*}[t]|\mb{Q}[t]=\mb{Q},\mb{T}[t]=\mb{T}],
\end{align}
where the step (a) is true for $\mb{r}^{*}=(r_l^{*})_{l=1}^{L}\in\argmax_{\mb{r}\in\mc{R}}\sum_{l=1}^{L}(\alpha_lQ_l[t]+\gamma\beta_lT_l[t])r_l$; step (b) follows from the fact that $\mb{r}^{*}\in\mc{R}$ and thus $\mb{r}^{*}$ can be written as $\mb{r}^{*}=\sum_{\mb{c}}\Pr\{\mb{C}[t]=\mb{c}\}\mb{v}^{(\mb{c})}$, where $\mb{v}^{(\mb{c})}\in\text{CH}\{\mc{S}^{(\mb{c})}\}$ for each channel state $\mb{c}$; and step (c) and (d) follow from equation \eqref{eqn:linear:prog} and equation \eqref{eqn:RSG:anotherform}, respectively.

By combing \eqref{eqn:pf:one} and \eqref{eqn:pf:second}, we have the desired result.

\begin{spacing}{}

\bibliographystyle{abbrv}
\bibliographystyle{IEEEtran}
\bibliography{refs}
\end{spacing}

\end{document}